\documentclass[aps,pre,showpacs,amsmath,amssymb,superscriptaddress,reprint,10pt,longbibliography,nofootinbib]{revtex4-1}
\usepackage{bm}
\usepackage{graphicx}
\usepackage{xcolor}
\usepackage{url}
\usepackage[normalem]{ulem}
\usepackage{times}
\usepackage{bbm}
\usepackage{mathrsfs} %Fancy letters 
\usepackage{comment}
\usepackage{dsfont}
\usepackage{multirow}
\usepackage{enumitem}
\usepackage{microtype}
\usepackage{siunitx}\DeclareSIUnit\gauss{G}
\usepackage[colorlinks=true,breaklinks=true,allcolors=blue]{hyperref}
\definecolor{philipp}{rgb}{1,.4,.3}
\definecolor{michael}{rgb}{0,.8,.5}
\definecolor{christian}{rgb}{.2,.6,1}

\newcommand\ZZ{{\mathds{Z}}}

\newcommand\CC{{\mathds{C}}}

\newcommand{\mvec}[1]{\boldsymbol #1}

\newcommand{\ket}[1]{ \displaystyle \left\vert #1 \right\rangle }
\newcommand{\bra}[1]{ \displaystyle \left\langle #1 \right\vert }
\newcommand{\expval}[1]{\left\langle #1 \right\rangle}
\newcommand{\matel}[3]{ \displaystyle \left\langle #1 \right \vert #2 \left\vert  #3 \right\rangle }

\DeclareMathOperator{\imp}{\text{Im}}
\DeclareMathOperator{\rep}{\text{Re}}

\DeclareMathOperator{\wmcf}{\mathscr{C}}
\DeclareMathOperator{\dist}{\text{dist}}

\begin{document}
	%
	%\defaultbibliography{LRLR}
	%\defaultbibliographystyle{plain}
	%
	\title{Probing unitary two-time correlations in a neutral atom quantum simulator}  
	\author{Philipp Uhrich} 
	%\affiliation{National Institute for Theoretical Physics (NITheP), Stellenbosch 7600, South Africa} 
	\affiliation{\mbox{Institute of Theoretical Physics,  Department of Physics, University of Stellenbosch, Stellenbosch 7600, South Africa}}
	\author{Christian Gross}
	\affiliation{Max-Planck-Institut f\"ur Quantenoptik, 85748 Garching, Germany}
	\author{Michael Kastner} 
	\email{kastner@sun.ac.za} 
	\affiliation{National Institute for Theoretical Physics (NITheP), Stellenbosch 7600, South Africa} 
	\affiliation{\mbox{Institute of Theoretical Physics,  Department of Physics, University of Stellenbosch, Stellenbosch 7600, South Africa}}
	\date{\today}
	\begin{abstract}
	Measuring unitarily-evolved quantum mechanical two-time correlations is challenging in general. In a recent paper [P.~Uhrich {\em et al.}, Phys.\ Rev.~A {\bf 96}, 022127 (2017)], a considerable simplification of this task has been pointed out to occur in spin-$1/2$ lattice models, bringing such measurements into reach of state-of-the-art or near-future quantum simulators of such models. Here we discuss the challenges of an experimental implementation of measurement schemes of two-time correlations in quantum gas microscopes or microtrap arrays. We propose a modified measurement protocol that mitigates these challenges, and we rigorously estimate the accuracy of the protocols by means of Lieb-Robinson bounds. On the basis of these bounds we identify a parameter regime in which the proposed protocols allow for accurate measurements of the desired two-time correlations.
	\end{abstract}
	%
	%\pacs{05.70.Ln, 05.20.-y, 05.30.-d, 05.50.+q} 
	%
	\maketitle 
	%------------------------------------------------
	%
	\section{Introduction}
	Over the past decade or so, quantum simulators, and in particular those based on cold atoms, molecules, or ions, have proved to be tremendously versatile: a large variety of different Hamiltonians has been realised, in one, two, or three spatial dimensions, in the continuum as well as on lattices of different structures; equilibrium as well as nonequilibrium physics has been probed; and a variety of observables has been measured, in some cases with single-atom spatial resolution and high temporal resolution~\cite{bloch2012, BlattRoos12, browaeys2016, GrossBloch17}. A quantity that so far has remained elusive in quantum simulation is temporal correlation functions of the type
	\begin{equation}\label{e:CO1O2}
	C(t_1,t_2):=\matel{\psi}{O_1(t_1)O_2(t_2)}{\psi},
	\end{equation}
	where
	\begin{equation}
	O_i(t_i)=e^{iHt_i}O_i e^{-iHt_i}
	\end{equation}
	denotes the observable $O_i$ in the Heisenberg picture, unitarily evolved under the Hamiltonian $H$ until time $t_i$. Such correlation functions are very popular with theoreticians, and feature prominently in many methods and theories in quantum dynamics. Examples include fluctuation-dissipation theorems and the Kubo formula \cite{Kubo57}, optical coherence \cite{Glauber63}, glassy dynamics and aging \cite{SciollaPolettiKollath15}, and many more.
	
	Such quantum mechanical two-time correlation functions are not easily accessible. The main reason is that a measurement at the earlier of the two times $t_1$ will in general strongly affect the state of the system and influence the outcome of the later measurement at time $t_2$. In short, measurement backaction destroys the unitarity of the quantum mechanical time evolution, and, being inherent to quantum mechanical measurements, makes experimental measurement of (1) extremely challenging. One possible way out is to devise schemes that give access to quantities encoding the two-time correlation function \eqref{e:CO1O2}, but that do not require any measurement at $t_1$. An example is linear response theory, which can give access to the imaginary part of $C$, but other, more specific schemes have been proposed as well \cite{RomeroIsart_etal12,Knap_etal13,YoshimuraFreericks}. The shortcoming of these schemes is that they work only in specific settings, for specific initial states, and/or give access only to certain specific correlation functions. Another strategy for measuring two-time correlation functions consists in reducing the measurement backaction by making use of weak measurements (or generalised measurements, or quantum measurements), as proposed in \cite{Uhrich_etal,KastnerUhrich}. While being applicable for very general Hamiltonians, initial states, and observables, the drawback of these protocols is that they require an exquisite control over the quantum system, and in particular the ability to temporarily couple auxiliary degrees of freedom to specific observables, as well as a large number of repetitions of the experiment in order to accumulate sufficient statistics.
	
	In the same Refs.~\cite{Uhrich_etal,KastnerUhrich}, a somewhat unexpected observation has been reported: for a certain class of Hamiltonians and observables, a projective measurement of the observable $O_1(t_1)$ at the earlier time $t_1$ has strictly no disturbing effect on the desired two-time correlation. This finding leads to a massive simplification compared to the above described weak-measurement protocol: no auxiliary degrees of freedom are required, and the required number of repetitions of the experiment is orders of magnitude smaller. In fact, the resulting measurement protocol is very simple: time-evolve the system until time $t_1$, measure $O_1$, time-evolve until $t_2$, and then measure $O_2$. As shown in Refs.~\cite{Uhrich_etal,KastnerUhrich}, this naive approach yields the real part of the desired two-time correlation function \eqref{e:CO1O2}, with strictly vanishing disturbance by measurement backaction, for arbitrarily-interacting systems of spin-$1/2$ degrees of freedom (or qubits) and single-site spin observables $\sigma_i^a$ (here, $\sigma$ denotes Pauli operators, $i$ refers to a single lattice site, and $a\in\{x,y,z\}$ denotes a spatial component). Similarly, and for the same class of Hamiltonians and observables, the imaginary part of the two-time correlation function \eqref{e:CO1O2} is obtained by replacing the measurement of $O_1$ at $t_1$ by a local unitary rotation. A detailed description of these ancilla-free measurement protocols (AFMPs) is given in Sec.~\ref{sec:prots} of the present paper and also in Refs.~\cite{Uhrich_etal,KastnerUhrich}.
	
	For these ancilla-free protocols to be implemented in a quantum simulator, the following is required: (a) emulation of a lattice spin-$1/2$ (qubit) model, (b) measurement of a spin component with single-site resolution, and (c) implementation of a single-site unitary rotation. All three ingredients are readily available in several of the existing platforms for quantum simulation, including neutral atom based platforms \cite{GrossBloch17} and ion traps \cite{BlattRoos12}. However, in both of these (and presumably in other) platforms an additional difficulty arises: the measurements affect the many-body state not only according to the von Neumann rule for projective measurements, but cause a lot more harm to the system. In fact, these disturbances can significantly alter the many-body spin state in the vicinity of the measured site, or even be so strong that atoms are lost after their measurement. One might think that such a strong disturbance---which we will subsequently refer to as \emph{measurement noise}---should have the potential to break the above described protocol entirely.
	
	The purpose of the present paper is to show that this is not true in general. By making use of Lieb-Robinson bounds \cite{LiebRobinson72,NachtergaeleSims10,KlieschGogolinEisert14}, we estimate the error induced by the abovementioned disturbances in the two-time correlation functions measured by means of the AFMPs. Denoting by $\rho$ the distance between the lattice sites at which $O_1$ and $O_2$ have nonvanishing support, and by $\Delta t=t_2-t_1$ the time between the two correlated events, we identify regions in the $(\rho, \Delta t)$-plane in which the effect of the measurement noise on the outcome of the two-time correlation function $C(t_1,t_2)$ is negligible and the AFMP remains valid. Such a disturbance due to measurement noise at $t_1$ is relevant only for the measurement protocol of the real part of the two-time correlation function.
	
	When measuring the imaginary part according to the AFMP, a local rotation is performed at $t_1$ instead of a projective measurement, and measurement noise therefore does not play a role. Other sources of noise may however still be present;  imprecisions of the rotation angle, for instance, may affect the accuracy of this protocol. Using similar, Lieb-Robinson-based tools, we show that the error induced by fluctuations in the rotation angle is negligible in similarly shaped regions in the $(\rho, \Delta t)$-plane in which the effect of the measurement noise was found to be negligible. Taken together, these results demonstrate that measurement imprecisions can be dealt with successfully in experimental implementations of the AFMPs, and both real and imaginary parts of two-time correlations $\eqref{e:CO1O2}$ are accessible at least in certain regions of the $(\rho, \Delta t)$-plane. While all these 
	results are rather general and of relevance for a number of quantum simulation platforms, we use the setting of a quantum gas microscope or microtrap array, and in particular the example of a long-range interacting spin lattice model emulated by means of Rydberg-dressed atoms, to illustrate our findings in Sec.~\ref{sec:rydberg}.
	
	In Sec.~\ref{sec:prots} we briefly summarize the ancilla-free measurement protocols (AFMP) for spin-$1/2$ lattices and single-site observables proposed in \cite{Uhrich_etal,KastnerUhrich}. In Sec.~\ref{sec:expimp} the potential implementation of the protocols with a neutral atom setup, and in particular the sources of imperfections (measurement noise and fluctuations in the rotation angle), are discussed. Secs.~\ref{sec:noiseRMP} and \ref{sec:noisePMP} contain the main theoretical results of this paper, namely bounds on the error that is introduced in the estimator of the two-time correlation function. The physical content of these bounds is illustrated in Sec.~\ref{sec:rydberg} for a model Hamiltonian that captures essential features of the long-range interacting spin model emulated by means of Rydberg-dressed atoms in the quantum gas microscope experiment of Ref.~\cite{Zeiher_etal2016}. A summary and a discussion of our main results is given in Sec.~\ref{sec:conc}.	
	\section{Measuring dynamic correlations of single-site spin-\texorpdfstring{$1/2$}{1/2} observables}
	\label{sec:prots}
	
	In this section we briefly summarize the ``ideal versions'' (without taking into account imperfections and measurement noise) of the ancilla-free measurement protocols put forward in \cite{Uhrich_etal,KastnerUhrich}. The protocols are applicable to arbitrary Hamiltonians $H$ of interacting spin-$1/2$ degrees of freedom (qubits), on arbitrary lattices, and for arbitrary initial states. The aim of the protocols is to extract real as well as imaginary parts of the two-time correlation function
	\begin{equation}\label{e:Csigmasigma}
	C(t_1,t_2):=\matel{\psi}{\sigma_i^a(t_1) O_2(t_2)}{\psi},
	\end{equation}
	where
	\begin{equation}
	\sigma_i^a(t)=U^\dagger(t)\sigma_i^a U(t)
	\end{equation}
	with
	\begin{equation}
	U(t)=\exp(-iHt).	
	\end{equation}
	$O_2$ is an arbitrary observable, which can be single-site or multi-site.
	\subsection{Rotation protocol for measuring \texorpdfstring{$\imp C$}{Im C}}
	\label{s:rp}
	The rotation protocol is based on the observation that the expectation value of $O_2$ with respect to the locally rotated state
	\begin{equation}\label{e:rotpsi}
	\ket{\psi(t_1,t_2,\theta)}=U(t_2-t_1)e^{-i \theta\sigma_i^a/2}U(t_1)\ket{\psi}
	\end{equation}
	is
	\begin{equation}\label{e:evRMP}
	\begin{split}
	E_{\theta} :=& \matel{\psi(t_1,t_2,\theta)}{O_2}{\psi(t_1,t_2,\theta)} \\
	=& \sin^2(\theta/2)\expval{\sigma_i^a(t_1) O_2(t_2) \sigma_i^a(t_1)}_\psi\\
	&+ \cos^2(\theta/2)\expval{O_2(t_2)}_\psi - \sin(\theta) \imp C(t_1,t_2),
	\end{split}
	\end{equation}
	where $\expval{\cdot}_\psi$ denotes an expectation value with respect to the initial state $\ket{\psi}$. The function $C$ in the last line of \eqref{e:evRMP} is the desired two-time correlation function \eqref{e:Csigmasigma}. $\imp C$ can be extracted from 
	\eqref{e:evRMP} by taking the difference
	\begin{equation}\label{e:diffRMP}
	\Delta E = E_{-\theta}-E_{\theta}= 2 \sin \theta \,\imp C(t_1,t_2)
	\end{equation}
	and dividing it by $2 \sin \theta$. The procedure for generating $\ket{\psi(t_1,t_2,\theta)}$ from an initial state $\ket{\psi}$ can be read off from the right-hand side of \eqref{e:rotpsi} and is summarised as the first step in the rotation protocol:
	\renewcommand{\labelenumi}{(\alph{enumi})}
	\begin{enumerate}[nosep]
		\item Initialise the system in the state $\ket{\psi}$, time-evolve it unitarily to $t_1$, apply the local rotation $e^{-i \theta\sigma_i^a/2}$ at site $i$, and then time-evolve until $t_2$. This yields $\ket{\psi(t_1,t_2,\theta)}$ as defined in \eqref{e:rotpsi}.
		\item Projectively measure $O_2$ and record the measured eigenvalue.
		\item Repeat steps (a) and (b) many times and use the relative frequencies with which each eigenvalue of $O_2$ is measured to estimate $E_\theta$.
		\item Repeat the above steps with rotation angle $-\theta$ in order to approximate $E_{-\theta}$. Use $E_{\theta}$ and $E_{-\theta}$ to calculate $\imp C(t_1,t_2)$ from \eqref{e:diffRMP}.
	\end{enumerate}
	As discussed in more detail in Refs.~\cite{Uhrich_etal,KastnerUhrich}, the statistical error in $\imp C$ caused by the finite number of repetitions of the measurement is minimised for the choice of $\theta = \pm \pi/2$.
	
	The rotation protocol resembles a linear response-type measurement scheme, but with the important difference that, for the class of systems and observables considered, it is valid not only in the linear regime of small $\theta$, but for arbitrary rotation angles.
	
	%--------------------------------------------------------
	%
	\subsection{Projective protocol for measuring \texorpdfstring{$\rep C$}{Re C}}
	\label{sec:protReC}
	Surprisingly, for spin-$1/2$ Hamiltonians and single-site observables, the naive approach of projectively measuring at time $t_1$ without worrying about measurement backaction, gives access to the real part of the desired correlation function \eqref{e:Csigmasigma}. This result should not be interpreted as arising from an absence of measurement backaction at $t_1$; wavefunction collapse \emph{does} occur, but, as \eqref{e:corrPMP} affirms, the effect of this collapse cancels out in the real part of correlations of single-site spin-$1/2$ observables. %A physically intuitive reason as to why this is the case, however remains elusive. 
	
	The projective protocol, originally put forward in \cite{Uhrich_etal,KastnerUhrich}, can be summarised as follows:
	\begin{enumerate}[nosep]
		\item Initialise the system in the state $\ket{\psi}$ and time-evolve it unitarily to $t_1$.
		\item Projectively measure $O_1=\sigma_i^a$ and record the measured eigenvalue $\pm 1$.
		\item Time evolve to $t_2$, measure $O_2$, and record the measured eigenvalue $\omega$.
		\item Repeat steps (a)--(c) many times so that the joint probabilities 
		$P^{\text{Proj}}(\pm,\omega)$ for obtaining the various combinations of outputs 
		of the two measurements can be estimated.
		\item Then, as shown in \cite{Uhrich_etal,KastnerUhrich}, the real part 
		of the two-time correlation function can be obtained from the equality 
			\begin{multline}\label{e:corrPMP}
		\rep\! \matel{\psi}{\sigma_i^a(t_1) 
			O_2(t_2)}{\psi}\\
			=\sum_{\omega}\omega\left(P^{\text{Proj}}(+,\omega)-P^{\text{Proj} }(-, \omega)\right),
		\end{multline}
		where the sum is over all eigenvalues $\omega$ of $O_2$.
	\end{enumerate}	
	This projective protocol can be used to measure $\rep \matel{\psi}{ O_1(t_1) O_2(t_2)}{\psi}$ when $O_1$ is not a single-site spin-$1/2$ operator as in \eqref{e:Csigmasigma}, but some other dichotomic observable
	\begin{equation}\label{e:dichot}
	O=e(\Pi_+-\Pi_-),
	\end{equation}
	where $\pm e$ are the two (possibly degenerate) eigenvalues of $O$, and $\Pi_\pm$ are the projections onto the corresponding eigenspaces. Examples of dichotomic observables are single-site spin-$1/2$ observables $\sigma_i^a$, or multi-site tensor products of these. Dichotomic observables exist also beyond spin-$1/2$ models, but are usually not of the form of experimentally accessible, local observables that are of physical relevance. For details on this see Sec.~$4.1$ of \cite{KastnerUhrich}. Results similar to those summarised in the present section have recently been published independently in Ref.~\cite{Dresseletal18}. 

	\section{Implementation in neutral atom quantum simulators}
	\label{sec:expimp}
	An implementation of the AFMPs of Secs.~\ref{s:rp} and \ref{sec:protReC} seems feasible in several of the experimental platforms that are currently being used as quantum simulators. Due the extraordinarily high level of control, trapped ions, quantum gas microscopes, or arrays of microtraps appear to be the most promising candidates. Here we discuss in detail strategies for the experimental implementation of AFMPs in the latter two, and the challenges that one has to expect. In these systems, ultra-cold atoms are trapped in a periodic light intensity distribution formed by an optical lattice or a microtrap array with atoms residing in the potential minima. This allows for high flexibility in the lattice geometries to be generated, and individual atoms can be addressed and manipulated. 
 	Making use of strongly interacting Rydberg atoms, light-induced interactions can be realised between the atoms, which allows for the emulation of a variety of spin-$1/2$ (and other) Hamiltonians in one- and two-dimensional lattices (for a review of these platforms see Refs.~\cite{browaeys2016,GrossBloch17} and references therein). The technique of Rydberg dressing used in Refs.~\cite{Zeiher_etal2016, zeiher2017a} allows one to tune the range and anisotropy of the simulated Ising interactions by off-resonantly coupling hyperfine ground states---used to simulate spin-$1/2$ particles---of the atoms to Rydberg states. 
  Importantly, Rydberg-dressing does neither compromise the optical trapping of the involved pseudo spin states, nor the possibility to perform local spin rotations. Furthermore,  various initial states can be created with high fidelity, giving access to equilibrium as well as nonequilibrium physics in such lattices.
	Proving the feasibility of our AFMPs on the neutral atom platform would thus allow dynamic correlations to be studied in various magnetic phases for a large class of magnetic Hamiltonians with various interaction ranges, such as the (an)isotropic Heisenberg spin models discussed in \cite{Glaetzle_etal2015, vanbijnen2015}.
	
	Implementation of the two AFMPs with neutral atoms, however, pose several severe challenges. Particularly, the rotation and projective measurement of site $i$ at $t_1$, required by the rotation protocol and projective protocol, respectively, are difficult to perform. This is true for existing Rydberg based platforms in optical lattices \cite{Zeiher_etal2016,guardado-sanchez2017} or arrays of optical microtraps \cite{labuhn2016, bernien2017}. In the remainder of this section we will motivate how these challenges can be circumvented through modifications of the AFMPs.
		
	We have seen in Sec.~\ref{sec:prots} that both AFMPs (for real and imaginary parts, respectively) require site- and spin-resolved projective measurements. These are readily available and have been used to observe site-resolved \emph{equal}-time spin and density correlations \cite{fukuhara2013a,Boll_etal16,Parsons_etal16,cheuk2016a,brown2017}. However, current implementations of site- and spin-resolved detection in these experiments are always global, and in most cases also destructive, in the sense that spin resolution is achieved by mapping one of the involved states to a lost atom after detection \cite{fukuhara2013a,Parsons_etal16, cheuk2016a,brown2017}. Even assuming that a projective detection at time $t_1$ could be performed in a local version of the routinely used global detection scheme, the detection would induce an unacceptably strong disturbance (beyond wave function collapse) of the spin lattice. Measuring only a single-site $i$ at $t_1$, as in the projective protocol, is thus extremely challenging on the commonly used alkali atom-based platform. The recently reported progress with alkaline-earth atoms in optical tweezers~\cite{cooper2018b,Norcia_etal} may allow one to use electron shelving techniques~\cite{nagourney1986,sauter1986,bergquist1986} to implement a local measurement in the future, which, however, we will not discuss here. Instead, we discuss a general detection scheme based on controlled relocation of the atoms to be measured. More straightforward is the realisation of a 
	controlled local rotation, as required at $t_1$ by the rotation protocol. In what follows we first discuss implementation of this rotation, before proceeding to the main part of this section where we propose a modified protocol that circumvents the difficulties, as outlined above, associated with local projective measurements at early times $t_1$. This modification is in principle applicable to quantum gas microscopes as well as microtrap arrays, but
	likely simpler to implement in the microtrap setting due to larger spatial separation of the traps.
	
	\subsection{Implementing a local rotation at \texorpdfstring{$t_1$}{t1}.}
	\label{sec:impRMP}
	
	The rotation protocol requires a site-resolved, controlled rotation of the $i$th qubit at $t_1$. This could be realized by a focused co-propagating Raman beam pair that is tuned in between the P$_{1/2}$ and P$_{3/2}$ states, such that the differential Stark shift (and thus any effective local magnetic field in the $\sigma^z$ direction) vanishes. To address only a single lattice site, the wavelength of these lasers should be on the order of the lattice spacing. Consequently, the pair of P states chosen to simulate the qubits must have an energy gap whose associated photon frequency is large enough to achieve the required spatial resolution. For example, in microtraps the lattice spacing (in the order of micrometers \cite{Barredo_etal16, endres2016}) allows one to choose wavelengths---and thus transitions---which belong to the typically visible or near-infrared transitions nS ~--~ nP. To cleanly select a single site in shorter-spaced optical lattices (with lattice constant of about $500\,\mathrm{nm}$) one might need to consider to use the nS~--~(n+1)P lines, which typically lie in the ultraviolet range for alkali atoms.
	
	One must further ensure that the remaining system experiences no dynamics while the $i$th qubit is rotated. In practice this means that the above rotation must be completed over a time $\tau$ which is much smaller than the shortest time-scale of the system Hamiltonian $H$. Fortunately, the Raman-coupling	 will allow for a fast spin rotation with duration $\tau \ll 1/J$, where $J$ is the largest pair-interaction energy in $H$, such that the evolution of the many-body system during the rotation is negligible. Alternatively, the laser induced interaction between the spins can be switched off during the rotation, simply by switching off the Rydberg dressing laser.
	Having implemented this localised, effectively instantaneous rotation of the $i$th qubit, the system can be let to evolve to $t_2$, at which point the final measurement can be performed with established global detection techniques, which provide the single-site resolution required to deduce the outcome of the measurement for the local observable $O_2$. An important consideration for the accuracy of the current AFMP is the effect of imperfect rotations of the $i$th qubit, which may be due to fluctuations in the Rabi frequency or due to magnetic field noise. For typical experimental conditions there is an uncertainty in the rotation angle on the order of a few percent. In Sec.~\ref{sec:noiseRMP} we analyse the impact of such noise on our measurement protocol.

	\subsection{Implementing a local projection at \texorpdfstring{$t_1$}{t1}.}
	\label{sec:impPMP}
\begin{figure}\centering
		\includegraphics[width=0.95\linewidth]{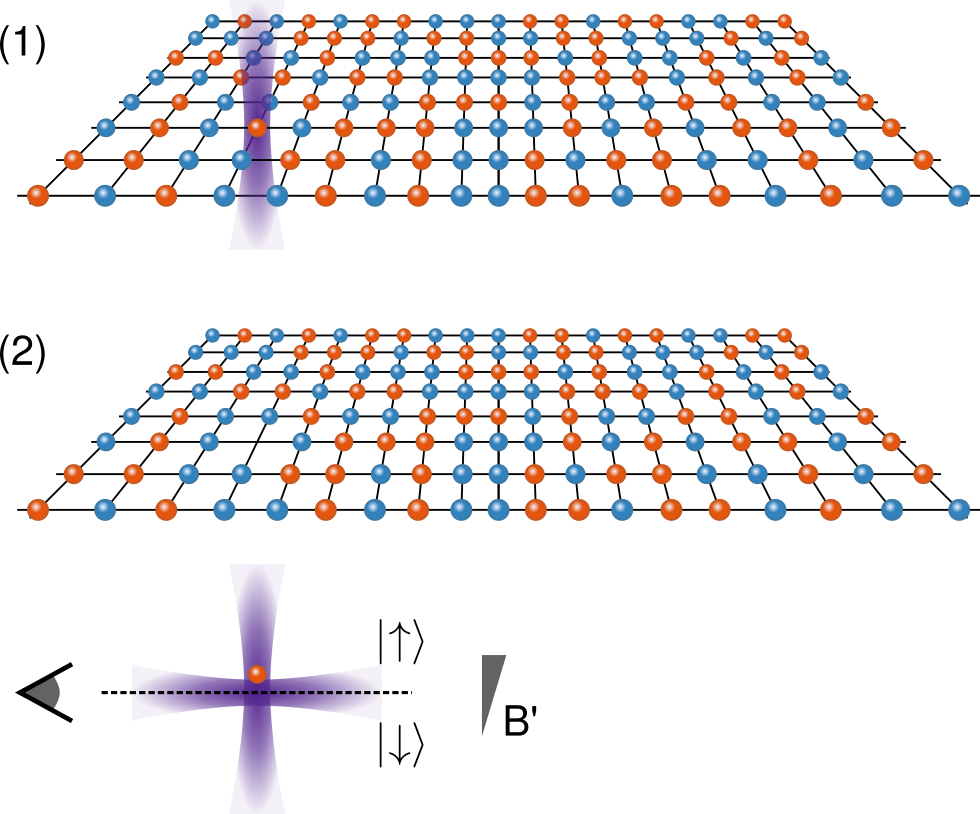}
 		\caption{\label{fig:scheme}
 			Schematic of the proposed projective measurement protocol. In the first step (1) a tightly focused optical tweezer selects the atom to be measured out of the lattice. The atom is then transported out of the atomic plane (2) and spin selectively measured from the side using a combination of magnetic field gradient $B'$ and repulsive light sheet (see text).
    }
 	\end{figure}

	The in-vivo measurement of $\rep C$ in the projective protocol requires a local 
	projective spin measurement at time $t_1$. Such a measurement is substantially more challenging than the local rotation described above, since the detection requires to scatter several hundreds to thousands of photons from a single atom and additionally involves several laser beams from different directions~\cite{endres2013,lester2014}. Under these conditions, it seems unfeasible to isolate the nearby lattice sites from the detection light.
  	As a possible solution, we propose to use a movable optical tweezer to transfer the spin of interest to a spatially well separated detection
	region displaced in the $z$ direction (see Fig.~\ref{fig:scheme}), i.e.\ below or above the atomic plane \cite{Barredo_etal16, barredo2017a}.
  An important assumption is that transferral of the $i$th atom to the detection region does not affect the many-body state $\ket{\psi(t_1))}$ of the lattice other than switching off all Hamiltonian terms involving the $i$th atom. This is assured by a tightly focussed tweezer, possibly with a wavelength near the UV/blue lines of the alkali atoms (the $nS - (n+1)P$ transitions), and the switching-off of the spin interactions during the measurement sequence as discussed in Sec.~\ref{sec:impRMP}.
  	After arrival in the detection region, the $i$th spin can be projectively measured in isolation (i.e.\ the remaining lattice is neither projected nor otherwise disturbed), thereby achieving the single-site projective measurement required by step $(b)$ of the projective protocol. The measurement of the in-plane spin components requires an additional rotation as in Sec.~\ref{sec:impRMP}, which should be implemented before moving the atom with the tweezer. In this way the spin direction of interest is rotated to the longitudinal readout direction, such that the protocol is immune to decoherence during the transport.
			
  	Spin resolution can be achieved by a Stern-Gerlach sequence similar to the technique demonstrated in Ref.~\cite{Boll_etal16}, where a magnetic field gradient normal to the atomic plane (parallel to the weak direction of the transport tweezer) is switched on. This gradient leads to a differential spatial displacement of the up ($\ket{+}$) and down ($\ket{-}$) spins when the internal states encoding the spin are chosen to have different magnetic moments.
  	Importantly, the gradient is normal to the atomic plane, such that it	induces only a trivial global phase evolution in the remaining spin system.
    If necessary, a spin echo can be used to cancel spurious differential phase shifts.
  	This spatial encoding of the spin state is required, since detection of the spin, e.g. by Raman sideband cooling-based imaging \cite{lester2014} mixes the hyperfine ground states. Furthermore, Raman sideband cooling often requires low magnetic fields. This can be achieved by focusing a blue-detuned light sheet in between the trap position for the up and down spins while the magnetic field is still on. Once the light sheet is placed, it defines two distinct traps for the different spins, even when the magnetic field is ramped down.
  	Finally, fluorescence is induced and the photons are observed from the side, such that the spatial position of the atom encoding the spin state can be resolved along the $z$ direction.

	Assuming that, at the end of the above described experimental implementation of a single-site projective measurement, the $i$th atom is lost, the system consists of one particle less, and evolves from time $t_1$ onwards under a modified Hamiltonian $H'$, which differs from the original Hamiltonian $H$ in that all terms involving atom $i$ have been removed. The projective correlation obtained under this modified Hamiltonian will, in general, differ from the correlation \eqref{e:corrPMP} obtained with the original projective protocol, i.e.\ an error is introduced which will affect the measurement of $O_2$ at $t_2$.
	The modified projective protocol is only useful if this error can be guaranteed to be small. In Sec.~\ref{sec:noisePMP} we provide a result that identifies rigorously the parameter regime for which this error is suppressed and the modified projective protocol can give a faithful estimate of the real part of the two-time correlation function.

	\section{Error propagation in the rotation protocol}
	\label{sec:noiseRMP}
	We have seen at the end of Sec.~\ref{sec:impRMP} that, when implementing the rotation protocol of Sec.~\ref{s:rp} in a quantum simulator, a possible, and presumably important, error source comes from imprecisions in the rotation angle $\theta$. The aim of the present section is to derive a bound on the final error that is induced by these fluctuations of $\theta$ in the measurements of the imaginary part of the two-time correlation $C$ defined in \eqref{e:Csigmasigma}. From Eq.~\eqref{e:diffRMP} we have that
	\begin{equation}\label{e:diff}
		\Delta E = E_{\theta_2}-E_{\theta_1} = 2\sin\theta\imp{C}
	\end{equation}
	for $\theta_1=-\theta_2=\theta$.
	To model an inaccuracy in the rotation angles, we parametrize their average deviation from the desired angle $\theta$ as
	\begin{equation}\label{e:thetamod}
		\theta_1=\theta+\delta_1 \qquad\theta_2=-\theta + \delta_2.
	\end{equation}
	The average deviations $\delta_1,\delta_2$ could be the result of a systematic inaccuracy of the experimental equipment, in which case $\delta_1,\delta_2$ would have the same value in every repetition of rotation protocol, i.e.\ $\delta_1,\delta_2=\delta$. A more likely source of error could be noise in the optical fields used to generate qubit rotations. This noise would vary in each run of the experiment, and follow a statistical distribution. Assuming this distribution to have zero mean and standard deviation $\sigma>0$, the resulting rotation angles will follow the same distribution, only with the mean shifted to $\pm \theta$. In this case $\delta_1,\delta_2=\sigma$. The following derivation is valid in either case, and we will continue to distinguish $\delta_1$ and $\delta_2$, to keep track of the errors incurred in the forward ($\theta_1$) and backward ($\theta_2$) rotations.
	
	We rewrite \eqref{e:evRMP} as
	\begin{equation}\label{e:eval}
		E_{\theta_m} = (\expval{1} -\expval{2})\cos^2(\theta_m /2) + \expval{2}  - \imp{C}\sin(\theta_m),
	\end{equation}
	where we have introduced the shorthand $\expval{1}\!=\!\expval{O_2(t_2)}_\psi$ and $\expval{2}\!=\!\expval{\sigma_i^a(t_1)O_2(t_2)\sigma_i^a(t_1)}_\psi$. Inserting the modified angles \eqref{e:thetamod} into \eqref{e:eval} we obtain %\pju{Check factor 4 here!}
	\begin{multline}\label{e:dE}
			\Delta E =  \left( \expval{1}-\expval{2} \right)\left[\cos^2\left(\frac{\theta-\delta_2}{2}\right) - \cos^2\left(\frac{\theta+\delta_1}{2}\right) \right]\\
			+\imp{C}\left(\sin(\theta\!-\!\delta_2)+\sin(\theta\!+\!\delta_1) \right) .
		\end{multline}
	Assuming the average deviation from the optimal rotation angles to be small, 
	we Taylor-expand \eqref{e:dE} with respect to $\delta_1,\delta_2$. For the optimal choice of rotation angle $\theta=\pi/2$, we obtain %\pju{Check factor 4 here! AND IN REMAINDER OF SECTION}
	\begin{equation}\label{e:dEfull}
		\Delta E = 2 \imp{C} + \epsilon ,
	\end{equation}
	with
	\begin{equation}\label{e:linEps}
		\epsilon = \frac{\delta_1+\delta_2}{2} (\expval{2}-\expval{1}) + \mathcal{O}(\delta_1^2) + \mathcal{O}(\delta_2^2).
	\end{equation}
	If the errors in the rotation angles are systematic, $\delta_1+\delta_2=2\delta$. If they are statistical, $\delta_1+\delta_2=2\sigma>0$. In either case the prefactor $\delta_1+\delta_2$ in \eqref{e:linEps} cannot vanish. To determine the size of the error \eqref{e:linEps}, one must therefore calculate the size of the scaling term
	\begin{equation}\label{e:e2-e1}
		\expval{2}-\expval{1} = \matel{\psi}{(\sigma_i^a(t_1)O_2(t_2)\sigma_i^a(t_1) - O_2(t_2))}{\psi} .
	\end{equation}
	In principle one could measure the two expectation values in \eqref{e:e2-e1} in additional experimental runs at an additional cost of time, money, and/or manpower. However, in order to be able to quantify the error $\epsilon$ in \eqref{e:linEps}, knowledge of the average deviations $\delta_1$ and $\delta_2$ would also be required. Here we avoid these difficulties by resorting to a theoretical analysis of \eqref{e:e2-e1} and identify conditions under which $\expval{2}-\expval{1}$, and hence $\epsilon$, is small. Our strategy is to upper bound $\lvert \expval{2}-\expval{1} \rvert$ by the operator norm of a suitable commutator, and then use a Lieb-Robinson bound to estimate that commutator. The advantage of such a bound is that it is independent of the (in general unknown) initial state $\ket\psi$ used in the rotation protocol of Sec.~\ref{s:rp}.
	
	The absolute value of \eqref{e:e2-e1} can be bounded by its operator norm,
	\begin{equation}
	\begin{split}
	\lvert \expval{2}-\expval{1} \rvert &\leq \lVert \sigma_i^a(t_1)O_2(t_2)\sigma_i^a(t_1) - O_2(t_2) \rVert\\
	&=\lVert \sigma_i^a(t_1) \left( O_2(t_2)\sigma_i^a(t_1) - \sigma_i^a(t_1)O_2(t_2) \right)\rVert.
	\end{split}
	\end{equation}
	Making use of ${\lVert A B \rVert \leq \lVert A \rVert \lVert B \rVert}$ and $\lVert \sigma_i^a(t_1) \rVert = 1 = \lVert U(t_1)\rVert$, we obtain
	\begin{equation}\label{e:2m1}
	\begin{split}
	\lvert \expval{2}-\expval{1} \rvert \leq& \lVert U^\dagger(t_1) \bigl( U(t_1) O_2(t_2) U^\dagger(t_1) \sigma_i^a \\
	&- \sigma_i^a U(t_1) O_2(t_2) U^\dagger(t_1) \bigr) U(t_1) \rVert \\
	\leq& \left\lVert \left[ O_2(\Delta t), \sigma_i^a \right] \right\rVert,
	\end{split}
	\end{equation}
	where $\Delta t =t_2 -t_1$ is the time between the rotation at $t_1$ and the measurement at $t_2$, during which errors in the rotation angle can propagate through the spin lattice. By these manipulations we have written the upper bound on $\lvert \expval{2}-\expval{1} \rvert$ in a form that can be further estimated by means of Lieb-Robinson bounds.
	
	Lieb and Robinson's original theorem \cite{LiebRobinson72}, as well as subsequent extensions to different settings \cite{Marchioro_etal78,HastingsKoma06,NachtergaeleSims06,BurrellOsborne07,Poulin10,MetivierBachelardKastner14,FossFeigGongClarkGorshkov15,StorchvandenWormKastner15,Matsuta2016,AbdulRahman_etal17}, give upper bounds on the operator norm of commutators of the form $[A(t),B]$,
	\begin{equation}\label{e:LR_general}
	\left\lVert[A(t),B]\right\rVert\leq b(t,x),
	\end{equation}
	where $A$ and $B$ are observables and $x=\dist(A,B)$ denotes the spatial separation\footnote{With respect to the 1-norm and in units of the lattice constant.} of $A=A(0)$ and $B$. In their original work \cite{LiebRobinson72}, which is valid for systems on regular lattices with finite-range interactions, Lieb and Robinson calculated a bound of the form
	\begin{equation}\label{e:LRoriginal}
	b_{\text{LR}}(t,x)=c\exp\left(\frac{vt-x}{\xi}\right),
	\end{equation}
	where the constants $c$, $v$, and $\xi$ depend on general features like lattice dimension and the interaction strength, but not on the details of the model. This bound tells us that there is a region in the $(t,x)$-plane, outside the cone defined by $vt\geq x$, where the norm of $[A(t),B]$ decays exponentially with the distance $x$. Other Lieb-Robinson-type bounds, like those for systems with long-range interactions \cite{HastingsKoma06,NachtergaeleSims06,MetivierBachelardKastner14,FossFeigGongClarkGorshkov15,StorchvandenWormKastner15}, may have a functional form different from \eqref{e:LRoriginal}, but they all have in common that they specify a certain (not necessarily cone-shaped) causal region, outside of which $\left\lVert[A(t),B]\right\rVert$ decays (although not necessarily exponentially).
	
	It is this causal behaviour which makes Lieb-Robinson bounds a useful tool for our error analysis:
	Substituting \eqref{e:2m1} and the Lieb-Robinson bound
	\begin{equation}\label{e:boundComm}
	\left\lVert \left[ O_2(\Delta t), \sigma_i^a \right] \right\rVert \leq b(\Delta t, \rho) \text{ with } \rho =d(\sigma_i^a,O_2)
	\end{equation}
	into \eqref{e:linEps}, we obtain the upper bound
	\begin{equation}\label{e:epsLR}
	\lvert \epsilon \rvert \leq \frac{\lvert \delta_1+\delta_2 \rvert}{2} b(\Delta t, \rho) + \mathcal{O}(\delta_1^2) + \mathcal{O}(\delta_2^2)
	\end{equation}
	on the error of $\Delta E$ in \eqref{e:dEfull}. This bound can be interpreted as follows: The average deviations $\delta_1,\delta_2$ of the angles in the rotation protocol can be thought of as unwanted perturbations of the system's state at time $t_1$. The bound \eqref{e:epsLR} specifies a region in the $(\Delta t, \rho)$ parameter space, outside of which the deviation of the measured value of the imaginary part \eqref{e:dEfull} of the correlation from its true value $\imp C$ is strongly suppressed. How strongly it is suppressed will depend on the specific Lieb-Robinson bound $b(t,x)$ applicable to the system under investigation. In many cases $b$ will be decaying exponentially with $x$ outside the causal region, in other cases the decay will be algebraic, depending on the type of interaction present in the Hamiltonian. The shape of the causal region will be illustrated in Sec.~\ref{sec:noisePMP} for the example of an Ising chain with long-range interactions, as it can be emulated in a quantum gas microscope by means of Rydberg dressing. For a given Hamiltonian $H$ we are thus able to determine a region in the $(\Delta t, \rho)$ plane for which the rotation protocol, despite inaccuracies $\delta_1$ and $\delta_2$ in the rotation angles, approximates $\imp C$ within a desired level of accuracy.
	\section{Error propagation in a modified projective protocol}
	\label{sec:noisePMP}
	In the ancilla-free measurement protocol for the real part of the two-time correlation function described in Sec.~\ref{sec:protReC}, a projective measurement of a single-site observable\footnote{Or of a multi-site observable satisfying \eqref{e:dichot}.} is required at the early time $t_1$. While experiments with quantum gas microscopes allow for measurements with single site resolution, these measurements (as discussed in 
	Sec.~\ref{sec:expimp}) disturb the system by more than just a local collapse of the wave function at the single site $i$ that is to be measured, which is detrimental to the correct performance of the proposed measurement protocol. A solution to this problem was outlined in Sec.~\ref{sec:expimp}, and it consists in removing the atom at site $i$ from the lattice and performing the measurement on the relocated atom. 
	In this case, for times $t>t_1$, i.e.\ after removing the atom from the lattice, the remainder of the system is described by a Hamiltonian $H'$ that differs locally from the original Hamiltonian $H$. For a general $N$-spin Hamiltonian with on-site and pair interactions,
	\begin{equation}\label{e:hqgm}
	H= \sum_{m=1}^N H_m  + \sum_{\substack{m,n=1\\ m \neq n}}^N 
	H_{mn},
	\end{equation}
	such a locally modified Hamiltonian is given by
	\begin{equation}\label{e:hmod}
	H'=H-\sum_{n\neq i}^N H_{in}= \sum_{m} 
	H_m + \sum_{\substack{m,n\neq i \\ m\neq n}} H_{mn},
	\end{equation}
	where all pair interactions terms involving lattice site $i$ have been eliminated.\footnote{Whether the on-site term $H_i$ is present or not in the decoupled Hamiltonian depends on the details of the experimental implementation, but the presence of this term is strictly irrelevant for what follows.} 

	In Sec.~\ref{s:modprot} we introduce a measurement protocol, similar to that of Sec.~\ref{sec:protReC}, that accounts for the situation of a modified time-evolution, governed by $H$ at times $t<t_1$, and by $H'$, at times $t>t_1$. The correlation function obtained in this way differs from the correlation of Sec.~\ref{sec:protReC}, and does not contain the desired real part of the unitarily evolved correlation function $C$. However, under suitable conditions, the two may be very close to each other, and we derive in Sec.~\ref{s:modproterror} a bound on their difference.
	This bound allows us to identify a region in parameter space in which the difference is small and the modified protocol can be used to faithfully measure the desired correlations. In Sec.~\ref{sec:rydberg} we apply the bound to the example of a long-range Hamiltonian as it can be emulated by means of Rydberg-dressed states in a neutral atom array.

	\subsection{Modified projective protocol}
	\label{s:modprot}
	We choose the correlated observables $O_1$ and $O_2$ in \eqref{e:CO1O2} to have 
	disjoint\footnote{The choice of initially disjoint supports ensures that none of the atoms in $X_2$, which will be probed at $t_2$, are removed from the system during the decoupling at $t_1$.} supports at $t=0$, i.e. $X_1 \cap X_2 = \varnothing$ with ${X_1=\text{supp}(O_1)}$ and ${X_2=\text{supp}(O_2)}$, and we denote the spectral representations of the two observables as
	\begin{equation}\label{e:AB}
	O_1=\sum_{\nu = \pm 1} \nu \Pi_{X_1}^{\nu},\qquad
	O_2=\sum_{\omega } \omega \Pi_{X_2}^{\omega} .
	\end{equation} 
	Here $\Pi_{X_1}^{\nu}$ is the projection operator of lattice region 
	$X_1$ corresponding to eigenvalue $\nu$ of $O_1$, and $\Pi_{X_2}^\omega$ is defined analogously for $O_2$. $O_1$ is chosen as a single-site observable $\sigma_i^a$, where $a\in\{x,y,z\}$ denotes an arbitrary spin component, and we have $X_1=\lbrace i \rbrace$. The modified correlation function $\wmcf^{\text{Proj}}_{H,H'}$ is defined as the outcome of the following protocol: 
	
	\paragraph{Initial state preparation and time evolution to $t_1$.}	 The initial state $\ket{\psi}$ is time-evolved until time $t_1$ under the dynamics generated by Hamiltonian $H$, leading to
	\begin{equation}\label{e:psi_t1}
	\ket{\psi(t_1)}=e^{-iHt_1}\ket{\psi}.
	\end{equation}
	
	\paragraph{Determining the post measurement state at $t_1$.} We assume that the removal of the $i$th atom from the lattice does not alter the many-body state of the system, but only its Hamiltonian, so $\ket{\psi(t_1)}$ given in \eqref{e:psi_t1} persists to be the state of the system by the time the removed atom $i$ is being measured, as assured by the procedure described in Sec.~\ref{sec:impPMP}. Under this assumption, and conditional on finding eigenvalue $\nu$ when measuring $O_1=\sigma_i^a$ at $t_1$, we obtain
	\begin{equation}\label{e:qgmPostMeas}
	%\begin{split}
	\ket{\psi_{\nu}(t_1)} = \Pi_{i}^{\nu} 
	e^{-iHt_1}\ket{\psi}/\sqrt{P^{\text{Proj}}_H(\nu)},
	%=&  \ket{\psi_{\pju{X_1}}(t_1)}\otimes \ket{\psi_{\pju{X_1^c}}(t_1)} 
	%\ps{\in \mathscr{H}_{X_1}\otimes\mathscr{H}_{X_1^c} = \mathscr{H}_S} ,
	%\end{split}
	\end{equation}
	where
	\begin{equation}
	P^{\text{Proj}}_H(\nu) = 
	\matel{\psi}{U^\dagger(t_1)\ket{\nu}\!\bra{\nu}U(t_1)}{\psi}
	\end{equation}
	is the probability of measuring eigenvalue $\nu$. The subscript $H$ indicates that this probability depends only on the 	original system Hamiltonian $H$, but not on $H'$. The post-measurement state $\ket{\psi_{\nu}(t_1)}$ is identical to the state obtained at the end of step (b) of the original protocol of Sec.~\ref{sec:protReC}.	
	
	\paragraph{Time-evolving to $t_2$ and measuring $O_2$.}
	With atom $i$ removed, the system evolves from $t_1$ to $t_2$ 	under the dynamics generated by the modified Hamiltonian $H'$.
	Then, as in the projective protocol of Sec.~\ref{sec:protReC}, observable $O_2$ is measured projectively at $t_2$. Using \eqref{e:qgmPostMeas} and \eqref{e:hmod} we find that the probability with which this measurement yields eigenvalue $\omega$, conditional on the outcome of the earlier measurement being $\nu$, is
	\begin{multline}\label{e:qgmCP}
	P^{\text{Proj}}_{H,H'}(\omega|\nu) = \matel{\psi}{ e^{i H t_1} 
		\Pi_{i}^{\nu} e^{i H'(t_2-t_1)}\Pi_{X_2}^{\omega} e^{-i H'(t_2-t_1)}\\
		\times \Pi_{i}^{\nu} e^{-i H t_1} }{\psi}/ 
	P^{\text{Proj}}_H(\nu) .
	\end{multline}
	Here we use the subscript $H,H'$ to indicate that \eqref{e:qgmCP} differs from the conditional probability of the original projective protocol due to the modified dynamics in the time interval $[t_1,t_2]$.
	
	\paragraph{Calculating correlations.}
	As in Sec.~\ref{sec:protReC}, we use the joint probabilities of obtaining the 
	various pairs of outcomes $(\nu,\omega)$ to calculate the modified projective 
	correlation as
	\begin{align}\label{e:cprop1}
	\wmcf_{H,H'}^{\text{Proj}} 
	=& \sum_{\nu =\pm 1, \omega} \nu \omega P^{\text{Proj}}_H(\nu) 
	P^{\text{Proj}}_{H,H'}(\omega \vert \nu) \nonumber\\
	%&=  \sum_{a} a \matel{\psi_0}{ e^{i H t_1} \Pi_{i}^{a} e^{i H' 
	%(t_2-t_1)} B e^{-i H' (t_2-t_1)} \Pi_{i}^{a} e^{-i H t_1}}{\psi_0} %\\
	=& \sum_{\nu =\pm 1} \nu \matel{\psi}{\Pi_{i}^{\nu}(t_1,H) 
		e^{i H t_1} O_2(\Delta t, H') e^{-i H t_1}\nonumber\\
		&\times \Pi_{i}^{\nu}(t_1,H) }{\psi} ,
	\end{align}
	where $\Delta t = t_2-t_1$ and we have used the spectral representations \eqref{e:AB} to obtain the second line of \eqref{e:cprop1}. To distinguish the Heisenberg time evolution of an operator $O$ under $H$ or $H'$ we have introduced the notation $O(t,H)=e^{iHt}Oe^{-iHt}$ and $O(t,H')=e^{iH't}Oe^{-iH't}$. Making use of spectral representation \eqref{e:AB} and the identity $\Pi_i^\pm=\mathds{1}-\Pi_i^\mp$, the modified projective correlation can be rewritten as
	\begin{multline}\label{e:cprop2}
	\wmcf_{H,H'}^{\text{Proj}} 
	= \matel{\psi}{\sigma_i^a(t_1,H) e^{i H t_1} O_2(\Delta t, H') e^{-i H t_1}}{\psi}\\
	\!+2i\imp\!\matel{\psi}{\Pi_{i}^-(t_1,H) 
		e^{i H t_1} O_2(\Delta t, H') e^{-i H t_1}\Pi_{i}^+(t_1,H)\!}{\psi}\!.
	\end{multline}
	Since $\wmcf_{H,H'}^{\text{Proj}}$ is real by definition, it follows that
	\begin{equation}\label{e:cprop3}
	\wmcf_{H,H'}^{\text{Proj}} 
	= \rep\matel{\psi}{\sigma_i^a(t_1,H) e^{i H t_1} O_2(\Delta t, H') e^{-i H t_1}}{\psi}.
	\end{equation}

	\subsection{Error of the modified protocol}
	\label{s:modproterror}
	
	The correlation function obtained by means of the original protocol of Sec.~\ref{sec:protReC}, where $H$ remains unchanged after the measurement at time $t_1$, is given by $\wmcf_{H}^{\text{Proj}}\equiv\wmcf_{H,H}^{\text{Proj}}$. We want to assess the accuracy of the modified projective protocol of Sec.~\ref{s:modprot} by comparing it to its original counterpart. The difference between the two is given by
	\begin{equation}\label{e:eps}
	\epsilon:=\left| \wmcf^{\text{Proj}}_{H} - \wmcf^{\text{Proj}}_{H,H'} \right|.
	\end{equation}
	Inserting \eqref{e:cprop3} and making use of the fact that $|\rep x|\leq|x|$ for all $x\in\CC$, we can bound the error as
	\begin{multline}\label{e:eps2}
	\epsilon\leq \bigl\lvert\bigl\langle\psi\big\vert\sigma_i^a(t_1,H)\\
	\times e^{iHt_1}\left[O_2(\Delta t,H)-O_2(\Delta t,H')\right]e^{-iHt_1}\big|\psi\bigr\rangle \bigr\rvert.
	\end{multline}
	We want to control the error for arbitrary (and in general unknown) initial states $\ket{\psi}$ and generic many-body Hamiltonians $H$, for which an exact evaluation of \eqref{e:eps} is in general out of reach. To eliminate the initial state dependence, we bound the expectation value in \eqref{e:eps2} by an operator norm,
	\begin{equation}\label{e:bound1}%\label{e:triangle}
		\epsilon \leq \left\lVert O_2(\Delta t,H)-O_2(\Delta t,H') \right\rVert,
	\end{equation}
	where we have used that $|\matel{\psi}{O}{\psi}|\leq \lVert O \rVert$ for any bounded operator $O$, the unitarity of time-evolution, and $\Vert\sigma_i^a\rVert=1$.
	
	\paragraph{A trivial bound on $\epsilon$}
	Making use of the triangle inequality as well as the unitarity of time evolution, \eqref{e:bound1} can be further simplified, resulting in the trivial bound
	\begin{equation}\label{e:trivial}
	\epsilon\leq\left\lVert O_2(\Delta t,H)\right\rVert+\left\lVert O_2(\Delta t,H') \right\rVert = 2\lVert O_2\rVert,
	\end{equation}
	which is constant in time and distance. This bound is never smaller than the desired two-time correlation function, so it will never give rise to a small {\em relative} error, and we have to work harder to come up with a stronger bound.
	
	Using Eq.~$(11)$ of Ref.~\cite{Kastner2015}, we can bound \eqref{e:bound1} by a temporal integral over norms of commutators. The error bound then becomes
	\begin{equation}\label{e:epsint}
	\begin{split}
	\epsilon \leq& \int_0^{\Delta t} d\tau \left\lVert \left[H-H', O_2(\tau, H') \right] \right\rVert \\
	\leq& \int_0^{\Delta t} d\tau \sum_{n\neq i} \left\lVert \left[H_{in}, O_2(\tau, H') \right] \right\rVert,
	\end{split}
	\end{equation}
	where we have used $H-H'=\sum_{n\neq i}H_{in}$.\footnote{Had we removed $H_i$ from $H$ in \eqref{e:hmod}, we would have $H-H'=H_i+\sum_{n\neq i}H_{in}$, which yields the same result \eqref{e:epsint} because then $[H_i, O_2(\tau,H')]=0$. This confirms our earlier statement that $\epsilon$ does not depend on the on-site Hamiltonian of the removed atom.} 
	We will now further bound the norm of the commutator in \eqref{e:epsint} in two different ways, and take the minimum of those two bounds and the trivial bound \eqref{e:trivial} as our final estimate of the error. 
	
	\paragraph{A slightly less trivial bound on $\epsilon$.}
	From a naive simplification of the norm of the commutator
	\begin{equation}
		\left\lVert \left[H_{in}, O_2(\tau, H') \right] \right\rVert \leq 2 \left\lVert H_{in} \right\rVert \left\lVert O_2 \right\rVert
	\end{equation}
	in \eqref{e:epsint} one obtains the error bound
	\begin{equation}\label{e:trivbound}
		\epsilon \leq E_{\text{naive}}(\Delta t) := 2 \Delta t 
		\left\lVert O_2 \right\rVert \sum_{n\neq i}\left\lVert H_{in} \right\rVert .
	\end{equation}
	However, by using this naive bound one loses all of the temporal physics captured by the commutators in \eqref{e:epsint}.
	
	\paragraph{Bounding $\epsilon$ with Lieb-Robinson bounds.}
	To retain the nontrivial time-dependence of \eqref{e:epsint}, we proceed by using Lieb-Robinson bounds $b(\tau,x)$ as defined in \eqref{e:LR_general} to approximate the size of the commutator norms as $ \left\lVert \left[H_{in}, O_2(\tau, H') \right] \right\rVert\leq b_{in}(\tau,x)$ where $x=\dist(H_{in}, O_2)$. Inserting this bound into \eqref{e:epsint}, the error can be estimated as
	\begin{equation}\label{e:nontrivbound}
		\epsilon \leq E(\Delta t, \rho) := \sum_{n\neq i} \int_0^{\Delta t} d\tau\, b_{in}(\tau, x),
	\end{equation}
	where $\rho$ is the shortest distance between lattice site $i$ and the support of $O_2$. The functional form of Lieb and Robinson's original bound \eqref{e:LRoriginal} and its generalisations \cite{Marchioro_etal78,HastingsKoma06,NachtergaeleSims06,BurrellOsborne07,Poulin10,MetivierBachelardKastner14,FossFeigGongClarkGorshkov15,StorchvandenWormKastner15,Matsuta2016,AbdulRahman_etal17}, and in particular the bounds' time dependence, is rather simple so that the temporal integral in \eqref{e:nontrivbound} can be easily performed.
	
	Lieb-Robinson bounds are usually optimised to capture the large-distance (large-$x$) behaviour, whereas they may grossly overshoot at short distance and late times [as in the original bound \eqref{e:LRoriginal} which grows exponentially with $t$]. As a result, the naive bound \eqref{e:trivbound}, and even the trivial one in \eqref{e:trivial}, may be smaller than the Lieb-Robinson bound \eqref{e:nontrivbound} for some range of $\Delta t$ and $\rho$, and we can optimise the final error bound on $\epsilon$ by combining all three of them,
	\begin{equation}\label{e:finalbound}
		\epsilon \leq \text{min}\left\{2\Vert O_2\rVert,E_{\text{naive}}(\Delta t), 
		E(\Delta t, \rho) \right\}.
	\end{equation}
	The bounds in Eqs.~\eqref{e:trivbound}--\eqref{e:finalbound} are the main result of Sec.~\ref{sec:noisePMP}, and they are general in that they apply to any spin-$1/2$ lattice Hamiltonian $H$, arbitrary initial states $\ket{\psi}$, and arbitrary observables $O_2$ whose support does not intersect that of $O_1= \sigma_i^a$.
	
	As in other applications of Lieb-Robinson bounds, the integrated bound of Eq.~\eqref{e:nontrivbound} defines a domain of $(\Delta t, \rho)$ coordinates---whose boundary depends on the type of interactions present in $H'$ \eqref{e:hmod}---outside of which $E(\Delta t, \rho)$, and hence $\epsilon$, is strongly suppressed [see the discussion below \eqref{e:LRoriginal}]. Hence, based on the bound \eqref{e:finalbound}, we can identify a region in the $(\rho,\Delta t)$ parameter space for which the error $\epsilon$ of the modified projective protocol of Sec.~\ref{s:modprot} is smaller than a chosen accuracy level. In this region, one can faithfully measure unitarily-evolved two-time correlation functions according to the proposed protocol.
	In the next section we will illustrate the size and shape of that region in parameter space for a concrete example.
	
	\subsection{Modified projective protocol implemented in a Rydberg-dressing based quantum simulator}
	\label{sec:rydberg}
	As an example we consider a spin-$1/2$ Ising Hamiltonian with long-range interactions that can be emulated by means of Rydberg dressing \cite{Zeiher_etal2016},
	\begin{equation}\label{e:hRyd}
	H= \sum_{m<n} U_{mn} \sigma_m^z \sigma_n^z + \sum_m \mvec{h}_m\cdot \mvec{\sigma}_m,
	\end{equation}
	where
	\begin{equation}
	U_{mn}=\frac{U_0}{1+(d(m,n)/R_c)^6}
	\end{equation}
	is the strength of the pair interactions in units where $\hbar =1$. 
	These interactions have a soft-core shape, saturating to $U_0$ when the distance between lattice sites $m$ and $n$ is smaller than the length scale $R_c$. For $d>R_c$, $U_{mn}$ decays rapidly, following an algebraic decay proportional to $d(m,n)^{-6}$ for large distances. The on-site terms in the second sum of \eqref{e:hRyd} arise from interactions between atoms and a site-dependent effective external field $\mvec{h}_m$ pointing (at least in principle) in an arbitrary spatial direction.
	
 	\begin{figure}\centering
		\includegraphics[width=0.85\linewidth]{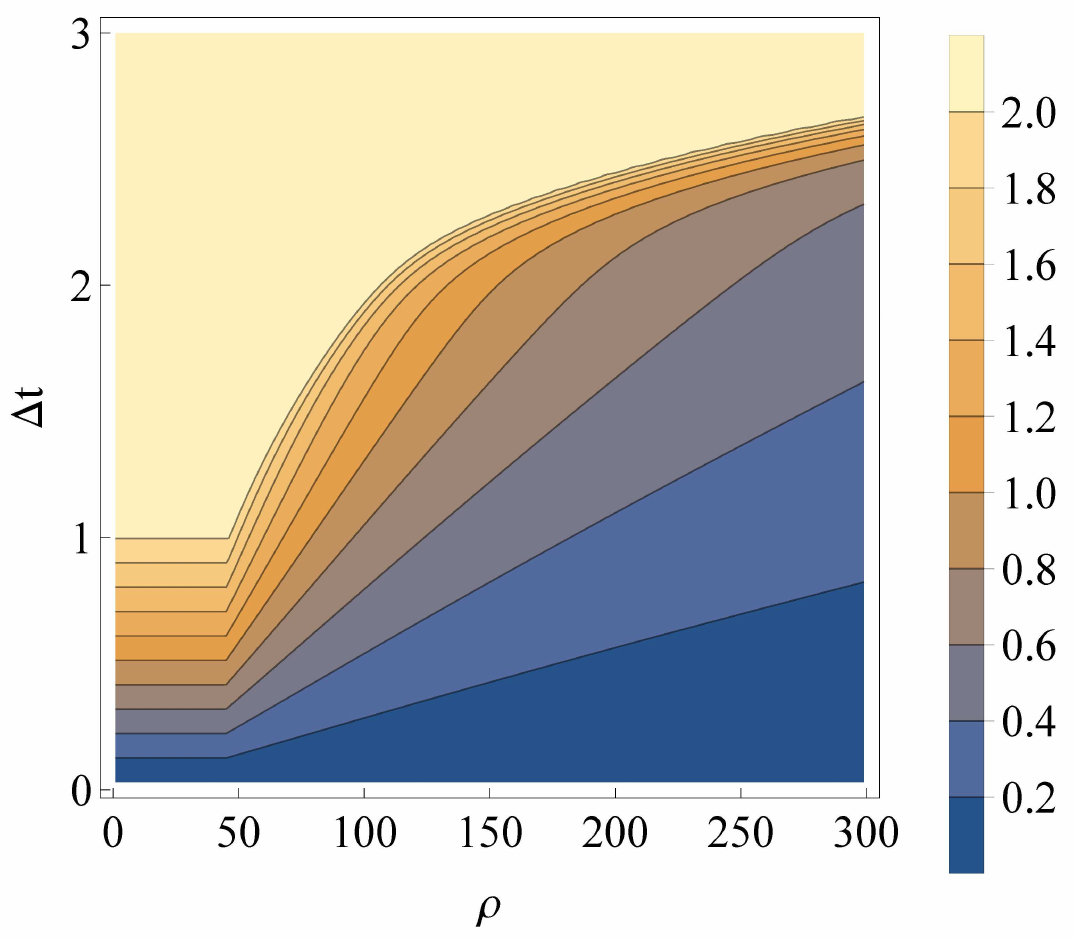}
		%\includegraphics[width=0.35\linewidth]{fig/bounds_legended_dt3_rho300.pdf}
		%\hspace{10mm}
		%\includegraphics[width=0.35\linewidth]{fig/logbounds_legended_dt3_rho300.pdf}
 		\caption{\label{fig:fullvstriv}%
 			Contour plot of the combined bound \eqref{e:finalbound} on the error $\epsilon$ for the long-range Ising Hamiltonian \eqref{e:hRyd} on a one-dimensional lattice. The quantity was evaluated numerically on the basis of \eqref{e:fullboundlong}. The parameters $U_0$ and $R_c$ in the Hamiltonian \eqref{e:hRyd} were set to one. For the exponents occurring in that bound, the long-range interactions $\propto\dist^{-6}$ imply $\kappa=1/3$ and $\eta=2/3$. Horizontal contours in the plot are a hallmark of regions where the $\rho$-independent naive bound $E_{\text{naive}}$ is tighter.
 			}
 	\end{figure}

	For the long-range interactions of \eqref{e:hRyd}, the appropriate choice for $b_{in}(\tau,x)$ in \eqref{e:nontrivbound} is the Lieb-Robinson bound \eqref{e:LRcube} of Matsuta {\em et al.} \cite{Matsuta2016}, which is a generalisation of the bound derived by Foss-Feig {\em et al.} \cite{FossFeigGongClarkGorshkov15}. Using this bound we determine an explicit form for $E(\Delta t, \rho)$ which can then be evaluated numerically. 
	We choose a single-site spin component $O_2=s_j^b$ and consider the Hamiltonian \eqref{e:hRyd} on a one-dimensional chain. The resulting error bound $E(\Delta t, \rho)$ is given by \eqref{e:fullboundlong} in Appendix~\ref{app:mats}, where also its functional dependence on $\Delta t=t_2-t_1$ and $\rho=d(O_1,O_2)=\vert i-j \vert$ is discussed. In Fig.~\ref{fig:fullvstriv} we plot the contours of \eqref{e:finalbound} in the $(\Delta t, \rho)$ parameter space, obtained by numerically evaluating \eqref{e:trivbound}~and~\eqref{e:fullboundlong}. For this calculation, the parameters $U_0$ and $R_c$ in the Hamiltonian \eqref{e:hRyd} have been set to 1, and hence the plot should be seen as a merely qualitative illustration of the features of the bound.  

	The horizontal sections of the contours in both plots in Fig.~\ref{fig:fullvstriv} are due to the $\rho$-independent naive bound $E_{\text{naive}}$ in \eqref{e:finalbound}. This confirms that indeed this bound is tighter than the Lieb-Robinson bound \eqref{e:nontrivbound} at short distances. The region below a chosen contour specifies the range of time differences $\Delta t$ and distances $\rho$ for which the modified projective protocol of Sec.~\ref{s:modprot} is guaranteed to faithfully reproduce the desired two-time correlation function $C$ up to a chosen accuracy level.

	\section{Conclusions}
	\label{sec:conc}

	Using as a starting point two recently proposed ancilla-free protocols for measuring unitarily-evolved two-time correlations $C(t_1,t_2)$ in spin-$1/2$ lattice systems \cite{Uhrich_etal}, we have discussed strategies for the implementation of such protocols in quantum gas microscopes or microtrap arrays. For an experimental implementation of the rotation protocol of Sec.~\ref{s:rp} for measuring the imaginary part of $C$, the main challenge is the level of accuracy with which a specific rotation angle can be realised. We were able to rigorously estimate, by means of Lieb-Robinson bounds, the size of the error that fluctuations of the rotation angle induce in the measured two-time correlations. The resulting bound \eqref{e:epsLR} is the first main result of the paper. On the basis of this bound, we can identify a region in the parameter space of time differences $\Delta t=t_2-t_1$ and distances $\rho=\dist(i,j)$ between lattice sites $i$ and $j$, such that the error induced by fluctuations of the rotation angle is smaller than some desired accuracy level $\epsilon$.
	
	For the projective protocol of Sec.~\ref{sec:protReC} that serves for measuring the real part of $C$, the main obstacle for an experimental implementation are the additional disturbances (beyond wave function collapse) that occur as a side effect of the fluorescence imaging that is employed for measuring the spin state at time $t_1$. We have proposed a modified projective protocol to mitigate these disturbances. The key idea here is to spatially separate the detection region from the physical system under study. Albeit challenging to implement, such a protocol is feasible with the state-of-the-art experimental techniques in quantum gas microscope and microtrap setups. This modified projective protocol is the second main result of the paper.
	
	The modifications made to the original projective protocol are unwanted from a theoretical perspective as, unlike in the original protocol, the resulting correlation function $\wmcf_{H,H'}^{\text{Proj}}$ of the modified protocol defined in Sec.~\ref{s:modprot} does not contain the correlation function $C$ which we are interested in. However, by resorting once more to Lieb-Robinson bounds, we are able to rigorously quantify the difference $\epsilon$ between $\rep C$ and the correlation function obtained from the modified projective protocol. This third main result of the paper is stated in Eqs.~\eqref{e:trivbound}--\eqref{e:finalbound}. These bounds allow us to identify a region in the parameter space $(\rho,\Delta t)$ where the modified projective protocol faithfully reproduces the desired real part of $C$ up to a chosen level of accuracy $\epsilon$. The thus obtained region of validity of the modified protocol will in general not exactly agree with that of the rotation protocol with angle fluctuations, but on a qualitative level the regions can be expected to be similar and have a significant overlap. In both cases, the region of validity will favour larger distances $\rho$ and shorter time differences $\Delta t$. 
	In particular, if the number of trapped particles does not allow for large separations $\rho$ (as is often case in quantum gas microscopes or ion trap experiments), our error bounds can be used to determine the allowed time interval $\Delta t$ for which the incurred error remains below a desired threshold.

	At this point it is worth noting that, although for the modified protocol to work the measurement at site $j$ and time $t_2$ has to be outside the ``causal region'' of the earlier measurement at site $i$ and time $t_1$, this does not imply that the resulting two-time correlation $C(t_1,t_2)$ is in any way trivial. This can be understood by noting that, after the removal and projective measurement of the atom at site $i$, the remainder of the system is in an, in general unknown and possibly strongly correlated, many-body state. Via these correlations the system's state, and also the outcome of the measurement at site $j$, at a later time $t_2$ will be affected in a nontrivial way by the entire lattice, including sites close to and beyond site $i$. In view of those, possibly strong and long-ranged, correlations it may be considered remarkable that the removal of the atom at site $i$ does not affect the accuracy of the modified projective protocol more strongly.

	Having understood the effect of the modification of the projective protocol on its accuracy, one may actually wonder whether other types of modifications might lead to better accuracy, i.e., to smaller errors $\epsilon$ than those we quantified in Eqs.~\eqref{e:trivbound}--\eqref{e:finalbound}. In Appendix~\ref{app:adap2} we present a variation of our modified protocol in which the $i$th atom is decoupled from the dynamics of the remaining lattice at $t_1$ as in the protocol of Sec.~\ref{sec:noisePMP}, but is only probed projectively at a later time $t_2$, simultaneous to the measurement of $O_2$. The idea behind this variation is that the decoupling preserves information about the $i$th atom's state at $t_1$, whilst the deferred measurement postpones any destructive effects of the measurement to the end of the experimental run. However, as shown by \eqref{e:boundprop2} in Appendix~\ref{app:adap2}, the error incurred in this way is, at least on the level of our bounds, as large or even larger than the error without deferral in \eqref{e:finalbound}, making the variation with deferral inferior to the modified projective protocol of Sec.~\ref{sec:noisePMP}.
  
 	The errors discussed here and throughout the paper are absolute errors as defined in \eqref{e:linEps} and \eqref{e:eps}. It would be desirable to also have bounds on relative errors, but, except for specific cases, this is unfortunately not a realistic goal. The problem is that, in order to derive an upper bound on the relative error, a {\em lower}\/ bound on the modulus of the actual value of the measured quantity is needed (in addition to the upper bound on the absolute error that we have). For a general initial state $\ket{\psi}$, the modulus of the two-time correlation $C(t_1,t_2)$ can be large or small (and even zero), irrespectively of the values of $\Delta t$ and $\rho$. Accordingly, an upper bound on the relative error is unrealistic for general initial states. What remains as a viable solution is to calculate relative errors on the basis of the actual measured values, combined with the bounds on absolute errors provided in this paper.

	As an outlook we would like to mention a potential strategy to further improve the accuracy of the experimental implementation of the projective protocol. The key idea here is to altogether avoid time-evolution by a locally modified Hamiltonian $H'$ [as defined in Eq.~\eqref{e:hmod}] by reinserting the projectively measured atom back into the lattice before proceeding with the time evolution up to time $t_2$. This approach requires, after the measurement, to transport the atom back precisely into its original position and that, after the complex measurement protocol, the temperature of the atom has not significantly increased. Both of these requirements can be realistically achieved with existing techniques: Precise relocation of atoms by moving tweezers has been demonstrated~\cite{Barredo_etal16} and optimized Raman sideband cooling allows to prepare the atom in a tweezer with high fidelity in its motional ground state~\cite{kaufman2012}.

	\begin{acknowledgments}
	P.U.\ acknowledges financial support from the National Research Foundation of South Africa, Stellenbosch University as well as the Sam Cohen Trust. C.G.\ acknowledges support from the European Research Council via the Starting Grant RyD-QMB and from Deutsche Forschungsgemeinschaft (Project No. GR 4741/1-1) within SPP 1929 (GiRyd)]. M.K.\ acknowledges financial support from the National Research Foundation of South Africa through the Competitive Programme for Rated Researchers. 
	\end{acknowledgments}	
	
	\appendix
	\section{Details of the Lieb-Robinson bounds used for Sec.~\ref{sec:rydberg}}
	\label{app:mats}
	Let $\Lambda$ denote the set of all lattice points in the spin-$1/2$ system. We make the following assumptions to simplify the subsequent derivation:
	\begin{enumerate}
		\item $\Lambda$ is a cubic lattice $\mathbb{Z}^D$,
		\item The lattice dimension is $D=1$ i.e.\ we consider a chain of qubits 
	\end{enumerate}
	For the long range interactions of \eqref{e:hRyd} we then have from Ref.~\cite{Matsuta2016} that $\lVert \left[ H_{in}, O_2(\tau, H') \right]\rVert $ is bounded from above by
	\begin{multline}\label{e:LRcube}
	b_{in}(\tau,x) = 2\lVert O_2 \rVert \lVert H_{in} 
	\rVert \lvert X_2 \rvert \Bigl( e^{v \tau - x/R} + 2 \tau g(x)f(R)  \\
	+ C_2 \lvert X_2 \rvert  R f(R) \tau e^{v \tau -x/R} \Bigr) ,
	\end{multline}
	with $x$ given by 
	\begin{align}\label{e:x}
	x=&d(H_{in}, O_2)=d(\{i,n\},\{j\})=\min \{d(j,i),d(j,n)\} \nonumber\\ 
	=& \min\{\rho,d(j,n)\}
	\end{align}
	and $X_2=\text{supp}(O_2)=j$. Bound \eqref{e:LRcube} is only valid for long-range systems in which the exponent $\alpha$ of the algebraically decaying interactions satisfies $\alpha > 2D$. In Hamiltonian \eqref{e:hRyd}, the interactions decay like $1/d(m,n)^6$ so that we have $\alpha=6$. Our derivation could therefore also be extended to lattices of dimension $D=2$ which are routinely implemented in quantum gas microscopes and microtrap arrays. In \eqref{e:LRcube} $R\geq 1$ is a length scale whose functional dependence on $x$ can be chosen so as to minimise the size of \eqref{e:LRcube} for specific regions of the $(\tau,x)$ parameter space [see \eqref{e:R}]. Parameters  $v$ and $C_2$ are positive constants independent of $\Lambda,\tau,X_1,X_2,O_2$ and $H_{in}$. Function $g(x)$ increases monotonically on $x\in [0,\infty )$ and is defined in Eq.~$(2.1)$ of \cite{Matsuta2016} as
	\begin{equation}\label{e:defg}
	\lvert \{m \in \Lambda \vert d(m,n) \leq x \} \rvert \leq g(x) \leq C(1+x)^D
	\end{equation}
	for some $C>0$, where $n\in \Lambda$ is some reference point. For $\Lambda=\ZZ$ we have
	\begin{equation}
		\lvert \{m \in \Lambda \vert d(m,n) \leq x \} \rvert \leq 2x< 2(1+x)
	\end{equation}
	and we thus use
	\begin{equation}
	g(x) = 2(1+x) \text{ for } x \geq 0
	\end{equation}
	in \eqref{e:fullboundlong} for our one-dimensional chain. 
	The function $f(R)$ in \eqref{e:LRcube} is a decreasing function on ${R\in [0,\infty )}$ and is defined in Assumption A of \cite{Matsuta2016} via
	\begin{equation}\label{e:deff}
	\sup_{m \in \Lambda} \sum_{ \substack{Z \ni m:\\ \text{diam}(Z) \geq R }} \lVert H_Z \rVert \leq f(R) .
	\end{equation}
	Here $H_Z$ is the interaction term of the Hamiltonian $H'$ which has support $Z \subset \Lambda$, and $\text{diam}(Z):=\text{max} \{ d(m,n)\vert m,n \in Z \}$. From \eqref{e:hRyd} we have
	\begin{equation}\label{e:hmodRyd}
		H'= \sum_{m=1}^N \mvec{h}_m \cdot \mvec{\sigma}_m + \sum_{\substack{m,n\neq i:\\ m< n}} U_{mn} \sigma_m^z \sigma_n^z 
%	\begin{split}
%		H'=& \sum_{m\neq i}^N h_m \pju{s_m^z} + \frac{1}{2}\sum_{\substack{m,n\neq i:\\ m\neq n}} U_{m,n} \pju{s_m^z s_n^z}  \\
%		=& \pju{H'_{X_1}\otimes \mathds{1}_{X^c_1} + \mathds{1}_{X_1} \otimes H'_{X^c_1}} ,
%	\end{split}
	\end{equation}
	from which follows that
	\begin{equation}
		H_Z  = \begin{cases}
		U_{mn} \sigma_m^z \sigma_n^z & \text{for $\lvert Z \rvert = 2$},\\
		0 & \text{otherwise}.
		\end{cases}
	\end{equation}
	Substituting this into the left-hand side of \eqref{e:deff} we have
	\begin{equation}\label{e:sup}
	\begin{split}
	%\sup_{x\in \Lambda}\sum_{\substack{Z\ni x:\\ \text{diam}(Z)\geq R}} \lVert h_Z \rVert =& 
	&\sup_{m\in \Lambda}\sum_{\substack{n\in \Lambda:\\ d(m,n)\geq R}} \lVert U_{mn} \sigma_m^z \sigma_n^z \rVert \\
	\leq& \sup_{m\in \Lambda}\sum_{\substack{n\in \Lambda:\\ d(m,n)\geq R}} U_0 (1+(d(m,n)/R_c)^6)^{-1}  \\
	\leq&  2 U_0 \sum_{d\geq R} (1+(d/R_c)^6)^{-1} \\
	\leq&  2U_0 \int_{R}^{\infty} dx \left[ 1+\left( \frac{x-1}{R_c} \right)^6 \right]^{-1} .
	\end{split}
	\end{equation}
	In the third line we assume that the system is large, and the resulting translational invariance allows us to drop the supremum. The factor $2$ arises because in $D=1$ dimensions there are always two lattice sites $n$ which are a distance $d$ away from a given site $m$. Bounding the sum in line three by the integral in the final line allows us to calculate a closed form of $f$ which depends only on $R$,
	\begin{equation}\label{e:f(R)}
	\begin{split}
	f(R) =& %2 U_0 \int_{R}^{\infty} dx (1+((x-1)/R_c)^6)^{-1} = 2 U_0 R_c \int_{\frac{R-1}{R_c}}^{\infty} \frac{du}{1+u^6} \\ =&
	\frac{2U_0 R_c}{6}\Biggl[2\pi + \arctan(\sqrt{3}-2\frac{R-1}{R_c})\\
	&- 2\arctan(\frac{R-1}{R_c})- \arctan(\sqrt{3}+2\frac{R-1}{R_c}) \\
	&-\sqrt{3}\,\text{arctanh}\left( \frac{\sqrt{3}\frac{R-1}{R_c}}{1+\left( \frac{R-1}{R_c}  \right)^2}\right)\Biggr] .
	\end{split}
	\end{equation}
	The parameter $v$ appearing in the exponentials of \eqref{e:LRcube} is defined in \cite{Matsuta2016} as
			\begin{equation}
			v=2eC_0 ,
			\end{equation}
			with the positive constant $C_0$ defined as
			\begin{equation}
			C_0 = \sup_{x \in \Lambda} \sum_{y\in \Lambda} \sum_{Z \ni x,y} \lVert H_Z \rVert < \infty .
			\end{equation}
			For a one-dimensional chain with Hamiltonian $H'$ \eqref{e:hmodRyd} one can show\footnote{Using the same logic as for the derivation of \eqref{e:f(R)}.} that $C_0=f(1)$, which implies
			\begin{equation}\label{e:v}
				v=2ef(1),
			\end{equation}
			and this is the value we used when evaluating the bound \eqref{e:fullboundlong} to produce the plots of Fig.~\ref{fig:fullvstriv}.
		The value of constant $C_2$ in \eqref{e:LRcube} can be inferred from Appendix~B of \cite{Matsuta2016} to be
	\begin{equation}
		C_2 = 8e^2.
\end{equation}
	
	Reference~\cite{Matsuta2016} shows that \eqref{e:LRcube} can be minimised by letting the length parameter $R$ scale with the distance $x$ defined in \eqref{e:x} as
	\begin{equation}\label{e:R}
	R = x^\kappa \text{ with } \kappa = (1+D)/(1+\alpha - D).
	\end{equation}
	Since $\alpha>2D$, $\kappa<1$ so that $R$ scales sub-linearly with $x$. Substituting \eqref{e:R} into \eqref{e:LRcube} we calculate the time integral of \eqref{e:LRcube},
	\begin{equation}\label{e:LRint}
	\int_0^{\Delta t} d\tau\, b_{in}(\tau, x) = U_{in} \beta(\Delta t, x)
	\end{equation}
	with
	\begin{multline}\label{e:beta}
	\beta(\Delta t, x) = 2 \Biggl( \frac{1}{v}\left( e^{v \Delta t } - 1\right)e^{- x^\eta} + 2 g(x) f(x^\kappa) \Delta t \\
	+ C_2 x^\kappa f(x^\kappa)\frac{1}{v^2}\left( e^{v \Delta t }\left( v\Delta t -1 \right) + 1 \right)e^{- x^\eta} \Biggr) .
	\end{multline}
	Here $\eta = 1-\kappa$, and we have used $\lvert X_2 \rvert = \lvert \{j\} \rvert  =1$, $\lVert O_2 \rVert = \lVert \sigma_j^b \rVert = 1$ and $\lVert H_{in}  \rVert= U_{in} \lVert  \sigma^z_i \sigma^z_n \rVert=U_{in}$. The reason for factoring the integrated bound in the above manner is that the spin-$1/2$ chain can be split into two disjoint domains according to the value taken on by $x$ as a function of $n$,
	\begin{equation}\label{e:domains}
	\begin{split}
	x=\rho \text{ for } n\in D_1&=\{n\in \Lambda\,|\, n\neq i \text{ and } d(j,n) \geq \rho \}\\
	x=d(j,n) \text{ for } n\in D_2&=\{n\in \Lambda\,|\, d(j,n)< \rho \},
	\end{split}
	\end{equation}
	with $\rho=d(j,i)$.	In $D_1$, $\beta(\Delta t, x)=\beta(\Delta t, \rho)$ is constant and we can split the summation of $E(\Delta t, \rho)$ \eqref{e:nontrivbound} as
	\begin{multline}\label{e:bound2}
		E(\Delta t, \rho)=  U_0 \biggl( \beta(\Delta t, \rho) \sum_{n\in D_1} \frac{1}{1+(d(i,n)/R_c)^6} \\
		+ \sum_{n\in D_2} \frac{\beta(\Delta t, x)}{1+(d(i,n)/R_c)^6}\biggr) .
		\end{multline}
	Since $D=1$, we have $\rho = |i-j|$ and $d(i,n)=|i-n|$ in units of the lattice constant. We further assume, without loss of generality, that $i<j$. Under this assumption domain $D_1$ lies to the left and right of $D_2$, and \eqref{e:bound2} can be written as
	\begin{multline}\label{e:fullboundshort}
		E(\Delta t, \rho)= U_0	\Bigl( \beta(\Delta t,\rho)  (\sum_{m=1}^{\infty}+\sum_{m=2\rho}^{\infty}) \frac{1}{1+(m/R_c)^6}\\
		+\sum_{m=1}^{2\rho -1} \frac{\beta(\Delta t, |		\rho - m|)}{1+(m/R_c)^6} \Bigr) ,
	\end{multline}
	where we have rewritten all distances occurring in the summations\footnote{{The first two summations in \eqref{e:fullboundshort} stem from the sum over $D_1$ in  \eqref{e:bound2} and the third summation is over $D_2$.}} in terms of ${m=|i-n| }$.
		
	Using \eqref{e:beta}, \eqref{e:fullboundshort} becomes
	\begin{widetext}
		\begin{equation}\label{e:fullboundlong}
		\begin{split}
		\frac{ E(\Delta t,\rho)}{ 2 U_0 } =& \frac{e^{v\Delta t}-1}{v} \biggl[ e^{-\rho^\eta} \left(s(1)+s(2\rho) \right) + \sum_{m=1}^\rho \frac{e^{-(\rho-m)^\eta}}{1+(m/R_c)^6} 
		+\sum_{m=\rho+1}^{2\rho-1} \frac{e^{-(m-\rho)^\eta}}{1+(m/R_c)^6}\biggr] + 2\Delta t \biggl[ g(\rho) f(\rho^\kappa) \left(s(1)+s(2\rho)\right)  \\
		&+ \sum_{m=1}^{\rho} \frac{g(\rho-m)f((\rho-m)^\kappa)}{1+(m/R_c)^6} +\sum_{m=\rho+1}^{2\rho-1}\frac{g(m-\rho)f((m-\rho)^\kappa)}{1+(m/R_c)^6}\biggr] 
		+ \frac{C_2}{v^2}\left( e^{v\Delta t}(v\Delta t -1) +1 \right) \biggl[ \rho^{\kappa} f(\rho^\kappa) e^{-\rho^\eta} \\
		&\times \left(s(1)+s(2\rho)\right)
		+ \sum_{m=1}^{\rho} \frac{(\rho -m)^{\kappa} f((\rho-m)^\kappa) e^{-(\rho-m)^\eta}}{1+(m/R_c)^6} +\sum_{m=\rho+1}^{2\rho-1} \frac{(m-\rho)^{\kappa} f((m-\rho)^\kappa) e^{-(m-\rho)^\eta}}{1+(m/R_c)^6} \biggr] .
		\end{split}
		\end{equation}
	\end{widetext}
	The physics of \eqref{e:fullboundlong} is to be understood as follows:
	The three square brackets correspond to the three terms of the temporally integrated Lieb-Robinson bound \eqref{e:LRint}. The temporal dependence is captured by the prefactors of the bracketed expressions. In the limit of large time intervals $\Delta t =t_2 -t_1$, these prefactors show that, for a fixed $\rho$, the overall error bound grows exponentially. The rate of this exponential growth is determined by the constant $v$ \eqref{e:v}.
	The spatial dependence of \eqref{e:fullboundlong} is more intricate and is captured by the terms within the square brackets. These terms stem from the spatial dependence of $\beta$ \eqref{e:LRint} and from the long-range interactions $U_{in}$ of $H_{in}$. The spatial contribution from $\beta$ is constant within domain $D_1$, and the corresponding summation, summarised in \eqref{e:fullboundlong} by $s(1)+s(2\rho)$ with
	\begin{equation}\label{e:sa}
	s(a) = \sum_{m=a}^{\infty} \frac{1}{1+(m/R_c)^6} ,
	\end{equation}
	depends only on the algebraic decay of the long-range interactions $U_{in}$. In \eqref{e:fullboundlong}, the explicitly shown sums are those which run over domain $D_2$ and they have been split into two physically distinct parts 
	\begin{equation}\label{e:subdomains}
	D_2 = d_1 \cup d_2
	\end{equation}
	with
	\begin{equation}
	d_1 =\{n\in D_2: i < n\leq j \},\quad d_2 =\{ n\in D_2: j<n \}
	\end{equation}
	satisfying $d_1\cap d_2 =\varnothing$.
	Within $d_1$, where $m=|i-n|\in[1,\rho]$, the spatial contributions of the long-range interactions $U_{in}$ and of $\beta$ compete: For lattice sites close to $i$ (i.e.\ summation index $m$ close to $1$) the algebraic terms $U_{in}$ are large and the exponential contribution from $\beta$ is small. As one moves further from $i$ and towards $j$, the algebraic interaction terms decay, but the exponential terms of $\beta$ grow. This exponential contribution is large for lattice sites close to $j$ ($m$ close to $\rho$). Sub-domain $d_2$ contains all lattice sites in $D_2$ which lie to the right of $j$, so that the summation runs over $m\in[\rho+1,2\rho-1]$. Within $d_2$, both the algebraic and the exponential terms decay as $m$ increases. This reflects the fact that one moves further from both lattice sites $i$ and $j$.

	The essential feature of \eqref{e:fullboundlong} is that at large times $\Delta t$ and distances $\rho$, $E(\Delta t, \rho)$ grows exponentially with $\Delta t$, but decays exponentially with $\rho$. 
	This competing exponential behaviour between these two spatial and temporal variables shows that the size of $E(\Delta t, \rho)$ can be minimised via optimal choices of measurement times $t_1$, $t_2$ and the separation $\rho$ of the supports of the correlated observables (see Fig.~\ref{fig:fullvstriv}).
	\section{Alternative modification of projective protocol with decoupling and deferral of measurements}
	\label{app:adap2}
	Our second adaptation of the projective protocol seeks to delay any destructive effects arising from the projective measurement of atoms in $\text{supp}(O_1)$ at $t_1$, to the final time $t_2$. For consistency we again choose $O_1=\sigma_i^a$, but we remind the reader that any observable of the form \eqref{e:dichot} is permissible. The idea is similar to the deferred measurement approach of Appendix~C of Ref.~\cite{Uhrich_etal}: At $t_1$ lattice site $i=\text{supp}(O_1)$ is not immediately measured (in contrast to the original projective protocol and the original adaptation of Sec.~\ref{sec:noisePMP}). Instead, the $i$th atom is decoupled from the remaining lattice at $t_1$ and its measurement is deferred to $t_2$,  at which time $X_2=\text{supp}(O_2)$ is probed simultaneously. Both projective measurements are thus performed at the final time, and any detrimental effects arising from these measurements are of no concern.
	
	To model the decoupling of site $i$ at $t_1$ we again let the lattice evolve from $t_1$ to $t_2$ under a decoupled Hamiltonian $H'$, the form of which is discussed after \eqref{e:cprop2}. Since no measurement is performed at the early times, the many-body state will in general not be a product state at $t_1$ [in contrast to \eqref{e:qgmPostMeas}]. Therefore, although site $i$ and the remaining lattice undergo decoupled dynamics under $H'$ for $t>t_1$ ($\tau > 0$), there is no guarantee that the $i$th spin evolves independently from the remaining system: Time evolution generated by $H$ during $0\leq t\leq t_1$ will in general entangle the state of the $i$th atom with the state of other atoms in the lattice. Dynamics of spins in $\Lambda \setminus \{i\}$ (where, as before, $\Lambda$ is the set of all lattice sites of the system), generated by $H'$ during $[t_1,t_2]$, will thus influence the $i$th spin, and vice versa. We therefore expect that the deviation $\epsilon$ of $\wmcf_{H,H'}^{\text{Proj}}$ from $\wmcf_{H}^{\text{Proj}}=\rep C$ will in general be larger than in the first adaptation of the projective protocol.
	
	The goal is again to bound $\epsilon=\lvert \wmcf_{H}^{\text{Proj}} - \wmcf_{H,H'}^{\text{Proj}}\rvert$ with Lieb-Robinson bounds, and much of the following error analysis is similar to that of Sec.~\ref{sec:noisePMP} so that we discuss only the pertinent differences to the derivation of \eqref{e:finalbound}.	
	Since site $i$ and region $X_2$ are now probed simultaneously at $t_2$, we deal with joint probabilities instead of conditional ones, and the probability of measuring eigenvalues $\nu$ and $\omega$, of $\sigma_i^a$ and $O_2$ respectively, is
	\begin{multline}\label{e:jointprob}
	P_{H,H'}^{\text{Proj}}(\nu, \omega) = \matel{\psi}{e^{i H 
			t_1}e^{iH'(t_2-t_1)} 
		\left( 
		\Pi_{i}^{\nu} \otimes \Pi_{X_2}^{\omega} \right) \\
		\times e^{-iH'(t_2-t_1)}e^{-iHt_1}}{\psi} .
	\end{multline}
	Using \eqref{e:jointprob}, the measured correlation is now
	\begin{equation}\label{e:cprop2prime}
	\begin{split}
	\wmcf_{H,H'}^{\text{Proj}} :=& \sum_{\nu=\pm 1, \omega } \nu \omega P_{H,H'}^{\text{Proj}}(\nu, \omega) \\
	=& \sum_{\nu=\pm 1} \nu 
	\matel{\psi}{e^{i H t_1}e^{i H' (t_2-t_1)} \left( \Pi_{i}^{\nu} \otimes O_2 
		\right)\\
		&\times e^{-i H' (t_2-t_1)} e^{-i H t_1}}{\psi} .
	\end{split}
	\end{equation}
	At this point the form of $H'$ must be considered: Our only requirement is that $H'$ does not contain any interactions $H_{mn}$ between spin $i$ and the remaining spins $n \neq i$. If one further removes the on-site energy term $H_i$, one obtains
	\begin{equation}\label{e:hmod2}
		H'=\mathds{1}_{i}\otimes H_{\Lambda\setminus \{i\}} .
	\end{equation}
	In this case $\left[ H', \Pi_{i}^{\nu} \right] = 0$, and \eqref{e:cprop2prime} is identical to the modified correlation \eqref{e:cprop1} of Sec.~\ref{sec:noisePMP}. The deviation from $\rep C$ is thus also bounded by \eqref{e:trivbound}--\eqref{e:finalbound} and we may conclude that when $H'$ is of the form \eqref{e:hmod2}, the deferred measurement approach presented here is as accurate as the modification of Sec.~\ref{sec:noisePMP}.

	In practice however, isolating the $i$th atom from the dynamics of the remaining system is achieved by locally increasing the trapping potential via a strong external magnetic field acting only on site $i$, and hence decoupling comes at the expense of a nonzero on-site Hamiltonian $H_i \neq 0$, such that
	\begin{equation}\label{e:hmod3}
	H'=H-\sum_{ n\neq i} H_{in} = H_{i}\otimes \mathds{1}_{\Lambda\setminus \{i\}} + \mathds{1}_{i} \otimes H_{\Lambda\setminus \{i\}}
	\end{equation}
	holds instead of \eqref{e:hmod2}.
	This modified Hamiltonian will in general yield a non-trivial commutator $\hat{c}(\nu, \Delta t):=\left[ e^{i\Delta t H'}, \Pi_{i}^{\nu} \right] =\left[ e^{i\Delta t H_{i}} , \Pi_{i}^{\nu} \right] \neq 0$, so that \eqref{e:cprop2prime} is not equal to \eqref{e:cprop1}, but rather given by (by the same logic which gave us \eqref{e:cprop3})
	\begin{multline}\label{e:cprop4}
		\wmcf_{H,H'}^{\text{Proj}} = \rep{\matel{\psi}{\sigma_i^a(t_1,H) e^{i H t_1} O_2(\Delta t, H') e^{-i H t_1}}{\psi}} \\
		+ \sum_{\nu=\pm 1} \nu \bra{\psi}e^{iHt_1} (\hat{c}O_2\hat{c}^\dagger \!+ \!(\hat{c}O_2e^{-iH'\Delta t} \Pi_i^\nu+ \text{h.c.})) \\
		 \times e^{-iHt_1}\ket{\psi} .
		\end{multline}	
	It follows that $\epsilon$ [obtained by substituting \eqref{e:corrPMP} and \eqref{e:cprop4} into \eqref{e:eps}] can be bounded by
	\begin{equation}\label{e:boundprop2}
	\begin{split}	
%	\epsilon \leq& \bigl\lvert\bigl\langle\psi\big\vert\sigma_i^a(t_1,H) e^{iHt_1}\left[O_2(\Delta t,H)-O_2(\Delta t,H')\right]e^{-iHt_1}\\
%	-& \!\sum_{\nu=\pm 1}\nu e^{iHt_1} ( \hat{c}O_2\hat{c}^\dagger \!+ \!(\hat{c}O_2e^{-iH'\Delta t} \Pi_i^\nu \!+\! \text{h.c.}) ) e^{-iHt_1}\big|\psi\bigr\rangle \bigr\rvert\\
	\epsilon \leq&  \lVert O_2(\Delta t, H) - O_2(\Delta t, H') \rVert \\&+ \sum_{\nu = \pm 1} \left( \lVert \hat{c} O_2 \hat{c}^\dagger \rVert + \lVert \Pi_{i}^{\nu} e^{i H' \Delta t} O_2 \hat{c}^\dagger + 
	\text{h.c.} \rVert \right) \\
	\leq& E'(\Delta t, \rho)
	\end{split}
	\end{equation}
	with
	\begin{multline}\label{e:E'}
	E'(\Delta t, \rho) := E(\Delta t, \rho) +\\
	\sum_{\nu = \pm 1} \left( \lVert \hat{c} O_2 \hat{c}^\dagger \rVert + \lVert \Pi_{i}^{\nu} e^{i H' \Delta t} O_2 \hat{c}^\dagger + \text{h.c.} \rVert \right) .
	\end{multline}
	The first line of \eqref{e:boundprop2} follows from the triangle inequality. The first term matches that of \eqref{e:bound1} and can therefore be approximated by $E(\Delta t, \rho)$ \eqref{e:nontrivbound}. The terms in the second line of \eqref{e:E'} imply that $E'(\Delta t, \rho) > E(\Delta t, \rho)$. Hence, at least on the level of our bounds, the protocol of Sec.~\ref{sec:noisePMP} approximates the desired two-time correlations $\rep C$ more accurately than the protocol with deferral.
	A possible reason why removal plus immediate measurement leads to a smaller error is that the projective measurement at the early time $t_1$ destroys any entanglement between the supports of the observables to be correlated, i.e.\ at $t_1$ the sites $i$ and $\Lambda\setminus \{i\} \supset X_2$ are decoupled. This also occurs in the original projective protocol, for which $\wmcf_H^{\text{Proj}}=\rep C$ exactly. In the  adaptation of this section however, the system's dynamics deviate more strongly from those of the original projective protocol, possibly because, in the absence of any measurement, lattice site $i$ and region $\Lambda\setminus \{i\}$ in general remain entangled for $t>t_1$.
	\bibliography{MyPub-modPMP}

\begin{thebibliography}{50}%
\makeatletter
\providecommand \@ifxundefined [1]{%
 \@ifx{#1\undefined}
}%
\providecommand \@ifnum [1]{%
 \ifnum #1\expandafter \@firstoftwo
 \else \expandafter \@secondoftwo
 \fi
}%
\providecommand \@ifx [1]{%
 \ifx #1\expandafter \@firstoftwo
 \else \expandafter \@secondoftwo
 \fi
}%
\providecommand \natexlab [1]{#1}%
\providecommand \enquote  [1]{``#1''}%
\providecommand \bibnamefont  [1]{#1}%
\providecommand \bibfnamefont [1]{#1}%
\providecommand \citenamefont [1]{#1}%
\providecommand \href@noop [0]{\@secondoftwo}%
\providecommand \href [0]{\begingroup \@sanitize@url \@href}%
\providecommand \@href[1]{\@@startlink{#1}\@@href}%
\providecommand \@@href[1]{\endgroup#1\@@endlink}%
\providecommand \@sanitize@url [0]{\catcode `\\12\catcode `\$12\catcode
  `\&12\catcode `\#12\catcode `\^12\catcode `\_12\catcode `\%12\relax}%
\providecommand \@@startlink[1]{}%
\providecommand \@@endlink[0]{}%
\providecommand \url  [0]{\begingroup\@sanitize@url \@url }%
\providecommand \@url [1]{\endgroup\@href {#1}{\urlprefix }}%
\providecommand \urlprefix  [0]{URL }%
\providecommand \Eprint [0]{\href }%
\providecommand \doibase [0]{http://dx.doi.org/}%
\providecommand \selectlanguage [0]{\@gobble}%
\providecommand \bibinfo  [0]{\@secondoftwo}%
\providecommand \bibfield  [0]{\@secondoftwo}%
\providecommand \translation [1]{[#1]}%
\providecommand \BibitemOpen [0]{}%
\providecommand \bibitemStop [0]{}%
\providecommand \bibitemNoStop [0]{.\EOS\space}%
\providecommand \EOS [0]{\spacefactor3000\relax}%
\providecommand \BibitemShut  [1]{\csname bibitem#1\endcsname}%
\let\auto@bib@innerbib\@empty
%</preamble>
\bibitem [{\citenamefont {Bloch}\ \emph {et~al.}(2012)\citenamefont {Bloch},
  \citenamefont {Dalibard},\ and\ \citenamefont {Nascimb\`ene}}]{bloch2012}%
  \BibitemOpen
  \bibfield  {author} {\bibinfo {author} {\bibfnamefont {I.}~\bibnamefont
  {Bloch}}, \bibinfo {author} {\bibfnamefont {J.}~\bibnamefont {Dalibard}}, \
  and\ \bibinfo {author} {\bibfnamefont {S.}~\bibnamefont {Nascimb\`ene}},\
  }\bibfield  {title} {\enquote {\bibinfo {title} {Quantum simulations with
  ultracold quantum gases},}\ }\href {\doibase 10/gcsjdc} {\bibfield  {journal}
  {\bibinfo  {journal} {Nat. Phys.}\ }\textbf {\bibinfo {volume} {8}},\
  \bibinfo {pages} {267--276} (\bibinfo {year} {2012})}\BibitemShut {NoStop}%
\bibitem [{\citenamefont {Blatt}\ and\ \citenamefont
  {Roos}(2012)}]{BlattRoos12}%
  \BibitemOpen
  \bibfield  {author} {\bibinfo {author} {\bibfnamefont {R.}~\bibnamefont
  {Blatt}}\ and\ \bibinfo {author} {\bibfnamefont {C.~F.}\ \bibnamefont
  {Roos}},\ }\bibfield  {title} {\enquote {\bibinfo {title} {Quantum
  simulations with trapped ions},}\ }\href {\doibase 10.1038/nphys2252}
  {\bibfield  {journal} {\bibinfo  {journal} {Nat. Phys.}\ }\textbf {\bibinfo
  {volume} {8}},\ \bibinfo {pages} {277--284} (\bibinfo {year}
  {2012})}\BibitemShut {NoStop}%
\bibitem [{\citenamefont {Browaeys}\ \emph {et~al.}(2016)\citenamefont
  {Browaeys}, \citenamefont {Barredo},\ and\ \citenamefont
  {Lahaye}}]{browaeys2016}%
  \BibitemOpen
  \bibfield  {author} {\bibinfo {author} {\bibfnamefont {A.}~\bibnamefont
  {Browaeys}}, \bibinfo {author} {\bibfnamefont {D.}~\bibnamefont {Barredo}}, \
  and\ \bibinfo {author} {\bibfnamefont {T.}~\bibnamefont {Lahaye}},\
  }\bibfield  {title} {\enquote {\bibinfo {title} {Experimental investigations
  of dipole{\textendash}dipole interactions between a few {Rydberg} atoms},}\
  }\href {\doibase 10/gddjbb} {\bibfield  {journal} {\bibinfo  {journal} {J.
  Phys. B}\ }\textbf {\bibinfo {volume} {49}},\ \bibinfo {pages} {152001}
  (\bibinfo {year} {2016})}\BibitemShut {NoStop}%
\bibitem [{\citenamefont {Gross}\ and\ \citenamefont
  {Bloch}(2017)}]{GrossBloch17}%
  \BibitemOpen
  \bibfield  {author} {\bibinfo {author} {\bibfnamefont {C.}~\bibnamefont
  {Gross}}\ and\ \bibinfo {author} {\bibfnamefont {I.}~\bibnamefont {Bloch}},\
  }\bibfield  {title} {\enquote {\bibinfo {title} {Quantum simulations with
  ultracold atoms in optical lattices},}\ }\href {\doibase
  10.1126/science.aal3837} {\bibfield  {journal} {\bibinfo  {journal}
  {Science}\ }\textbf {\bibinfo {volume} {357}},\ \bibinfo {pages} {995--1001}
  (\bibinfo {year} {2017})}\BibitemShut {NoStop}%
\bibitem [{\citenamefont {Kubo}(1957)}]{Kubo57}%
  \BibitemOpen
  \bibfield  {author} {\bibinfo {author} {\bibfnamefont {R.}~\bibnamefont
  {Kubo}},\ }\bibfield  {title} {\enquote {\bibinfo {title}
  {Statistical-mechanical theory of irreversible processes. {I}. {G}eneral
  theory and simple applications to magnetic and conduction problems},}\
  }\href@noop {} {\bibfield  {journal} {\bibinfo  {journal} {J. Phys. Soc.
  Jpn.}\ }\textbf {\bibinfo {volume} {12}},\ \bibinfo {pages} {570--586}
  (\bibinfo {year} {1957})}\BibitemShut {NoStop}%
\bibitem [{\citenamefont {Glauber}(1963)}]{Glauber63}%
  \BibitemOpen
  \bibfield  {author} {\bibinfo {author} {\bibfnamefont {R.~J.}\ \bibnamefont
  {Glauber}},\ }\bibfield  {title} {\enquote {\bibinfo {title} {The quantum
  theory of optical coherence},}\ }\href {\doibase 10.1103/PhysRev.130.2529}
  {\bibfield  {journal} {\bibinfo  {journal} {Phys. Rev.}\ }\textbf {\bibinfo
  {volume} {130}},\ \bibinfo {pages} {2529--2539} (\bibinfo {year}
  {1963})}\BibitemShut {NoStop}%
\bibitem [{\citenamefont {Sciolla}\ \emph {et~al.}(2015)\citenamefont
  {Sciolla}, \citenamefont {Poletti},\ and\ \citenamefont
  {Kollath}}]{SciollaPolettiKollath15}%
  \BibitemOpen
  \bibfield  {author} {\bibinfo {author} {\bibfnamefont {B.}~\bibnamefont
  {Sciolla}}, \bibinfo {author} {\bibfnamefont {D.}~\bibnamefont {Poletti}}, \
  and\ \bibinfo {author} {\bibfnamefont {C.}~\bibnamefont {Kollath}},\
  }\bibfield  {title} {\enquote {\bibinfo {title} {Two-time correlations
  probing the dynamics of dissipative many-body quantum systems: Aging and fast
  relaxation},}\ }\href {\doibase 10.1103/PhysRevLett.114.170401} {\bibfield
  {journal} {\bibinfo  {journal} {Phys. Rev. Lett.}\ }\textbf {\bibinfo
  {volume} {114}},\ \bibinfo {pages} {170401} (\bibinfo {year}
  {2015})}\BibitemShut {NoStop}%
\bibitem [{\citenamefont {Romero-Isart}\ \emph {et~al.}(2012)\citenamefont
  {Romero-Isart}, \citenamefont {Rizzi}, \citenamefont {Muschik}, \citenamefont
  {Polzik}, \citenamefont {Lewenstein},\ and\ \citenamefont
  {Sanpera}}]{RomeroIsart_etal12}%
  \BibitemOpen
  \bibfield  {author} {\bibinfo {author} {\bibfnamefont {O.}~\bibnamefont
  {Romero-Isart}}, \bibinfo {author} {\bibfnamefont {M.}~\bibnamefont {Rizzi}},
  \bibinfo {author} {\bibfnamefont {C.~A.}\ \bibnamefont {Muschik}}, \bibinfo
  {author} {\bibfnamefont {E.~S.}\ \bibnamefont {Polzik}}, \bibinfo {author}
  {\bibfnamefont {M.}~\bibnamefont {Lewenstein}}, \ and\ \bibinfo {author}
  {\bibfnamefont {A.}~\bibnamefont {Sanpera}},\ }\bibfield  {title} {\enquote
  {\bibinfo {title} {Quantum memory assisted probing of dynamical spin
  correlations},}\ }\href {\doibase 10.1103/PhysRevLett.108.065302} {\bibfield
  {journal} {\bibinfo  {journal} {Phys. Rev. Lett.}\ }\textbf {\bibinfo
  {volume} {108}},\ \bibinfo {pages} {065302} (\bibinfo {year}
  {2012})}\BibitemShut {NoStop}%
\bibitem [{\citenamefont {Knap}\ \emph {et~al.}(2013)\citenamefont {Knap},
  \citenamefont {Kantian}, \citenamefont {Giamarchi}, \citenamefont {Bloch},
  \citenamefont {Lukin},\ and\ \citenamefont {Demler}}]{Knap_etal13}%
  \BibitemOpen
  \bibfield  {author} {\bibinfo {author} {\bibfnamefont {M.}~\bibnamefont
  {Knap}}, \bibinfo {author} {\bibfnamefont {A.}~\bibnamefont {Kantian}},
  \bibinfo {author} {\bibfnamefont {T.}~\bibnamefont {Giamarchi}}, \bibinfo
  {author} {\bibfnamefont {I.}~\bibnamefont {Bloch}}, \bibinfo {author}
  {\bibfnamefont {M.~D.}\ \bibnamefont {Lukin}}, \ and\ \bibinfo {author}
  {\bibfnamefont {E.}~\bibnamefont {Demler}},\ }\bibfield  {title} {\enquote
  {\bibinfo {title} {Probing real-space and time-resolved correlation functions
  with many-body {R}amsey interferometry},}\ }\href {\doibase
  10.1103/PhysRevLett.111.147205} {\bibfield  {journal} {\bibinfo  {journal}
  {Phys. Rev. Lett.}\ }\textbf {\bibinfo {volume} {111}},\ \bibinfo {pages}
  {147205} (\bibinfo {year} {2013})}\BibitemShut {NoStop}%
\bibitem [{\citenamefont {Yoshimura}\ and\ \citenamefont
  {Freericks}(2016)}]{YoshimuraFreericks}%
  \BibitemOpen
  \bibfield  {author} {\bibinfo {author} {\bibfnamefont {B.~T.}\ \bibnamefont
  {Yoshimura}}\ and\ \bibinfo {author} {\bibfnamefont {J.~K.}\ \bibnamefont
  {Freericks}},\ }\bibfield  {title} {\enquote {\bibinfo {title} {Measuring
  nonequilibrium retarded spin-spin {G}reen's functions in an ion-trap-based
  quantum simulator},}\ }\href {\doibase 10.1103/PhysRevA.93.052314} {\bibfield
   {journal} {\bibinfo  {journal} {Phys. Rev. A}\ }\textbf {\bibinfo {volume}
  {93}},\ \bibinfo {pages} {052314} (\bibinfo {year} {2016})}\BibitemShut
  {NoStop}%
\bibitem [{\citenamefont {Uhrich}\ \emph {et~al.}(2017)\citenamefont {Uhrich},
  \citenamefont {Castrignano}, \citenamefont {Uys},\ and\ \citenamefont
  {Kastner}}]{Uhrich_etal}%
  \BibitemOpen
  \bibfield  {author} {\bibinfo {author} {\bibfnamefont {P.}~\bibnamefont
  {Uhrich}}, \bibinfo {author} {\bibfnamefont {S.}~\bibnamefont {Castrignano}},
  \bibinfo {author} {\bibfnamefont {H.}~\bibnamefont {Uys}}, \ and\ \bibinfo
  {author} {\bibfnamefont {M.}~\bibnamefont {Kastner}},\ }\bibfield  {title}
  {\enquote {\bibinfo {title} {Noninvasive measurement of dynamic correlation
  functions},}\ }\href {\doibase 10.1103/PhysRevA.96.022127} {\bibfield
  {journal} {\bibinfo  {journal} {Phys. Rev. A}\ }\textbf {\bibinfo {volume}
  {96}},\ \bibinfo {pages} {022127} (\bibinfo {year} {2017})}\BibitemShut
  {NoStop}%
\bibitem [{\citenamefont {Kastner}\ and\ \citenamefont
  {Uhrich}(2018)}]{KastnerUhrich}%
  \BibitemOpen
  \bibfield  {author} {\bibinfo {author} {\bibfnamefont {M.}~\bibnamefont
  {Kastner}}\ and\ \bibinfo {author} {\bibfnamefont {P.}~\bibnamefont
  {Uhrich}},\ }\bibfield  {title} {\enquote {\bibinfo {title} {Reducing
  backaction when measuring temporal correlations in quantum systems},}\ }\href
  {\doibase 10.1140/epjst/e2018-00086-8} {\bibfield  {journal} {\bibinfo
  {journal} {Eur. Phys. J. Special Topics}\ }\textbf {\bibinfo {volume}
  {227}},\ \bibinfo {pages} {365--378} (\bibinfo {year} {2018})}\BibitemShut
  {NoStop}%
\bibitem [{\citenamefont {Lieb}\ and\ \citenamefont
  {Robinson}(1972)}]{LiebRobinson72}%
  \BibitemOpen
  \bibfield  {author} {\bibinfo {author} {\bibfnamefont {E.~H.}\ \bibnamefont
  {Lieb}}\ and\ \bibinfo {author} {\bibfnamefont {D.~W.}\ \bibnamefont
  {Robinson}},\ }\bibfield  {title} {\enquote {\bibinfo {title} {The finite
  group velocity of quantum spin systems},}\ }\href {\doibase
  10.1007/BF01645779} {\bibfield  {journal} {\bibinfo  {journal} {Commun. Math.
  Phys}\ }\textbf {\bibinfo {volume} {28}},\ \bibinfo {pages} {251--257}
  (\bibinfo {year} {1972})}\BibitemShut {NoStop}%
\bibitem [{\citenamefont {Nachtergaele}\ and\ \citenamefont
  {Sims}(2010)}]{NachtergaeleSims10}%
  \BibitemOpen
  \bibfield  {author} {\bibinfo {author} {\bibfnamefont {B.}~\bibnamefont
  {Nachtergaele}}\ and\ \bibinfo {author} {\bibfnamefont {R.}~\bibnamefont
  {Sims}},\ }\bibfield  {title} {\enquote {\bibinfo {title} {{L}ieb-{R}obinson
  bounds in quantum many-body physics},}\ }in\ \href@noop {} {\emph {\bibinfo
  {booktitle} {Entropy and the Quantum}}},\ \bibinfo {series} {{Contemporary
  Mathematics}}, Vol.\ \bibinfo {volume} {529},\ \bibinfo {editor} {edited by\
  \bibinfo {editor} {\bibfnamefont {R.}~\bibnamefont {Sims}}\ and\ \bibinfo
  {editor} {\bibfnamefont {D.}~\bibnamefont {Ueltschi}}}\ (\bibinfo
  {publisher} {American Mathematical Society, Providence},\ \bibinfo {year}
  {2010})\BibitemShut {NoStop}%
\bibitem [{\citenamefont {Kliesch}\ \emph {et~al.}(2014)\citenamefont
  {Kliesch}, \citenamefont {Gogolin},\ and\ \citenamefont
  {Eisert}}]{KlieschGogolinEisert14}%
  \BibitemOpen
  \bibfield  {author} {\bibinfo {author} {\bibfnamefont {M.}~\bibnamefont
  {Kliesch}}, \bibinfo {author} {\bibfnamefont {C.}~\bibnamefont {Gogolin}}, \
  and\ \bibinfo {author} {\bibfnamefont {J.}~\bibnamefont {Eisert}},\
  }\bibfield  {title} {\enquote {\bibinfo {title} {{L}ieb-{R}obinson bounds and
  the simulation of time-evolution of local observables in lattice systems},}\
  }in\ \href {\doibase 10.1007/978-3-319-06379-9_17} {\emph {\bibinfo
  {booktitle} {Many-Electron approaches in Physics, Chemistry and
  Mathematics}}},\ \bibinfo {editor} {edited by\ \bibinfo {editor}
  {\bibfnamefont {L.~D.}\ \bibnamefont {Site}}\ and\ \bibinfo {editor}
  {\bibfnamefont {V.}~\bibnamefont {Bach}}}\ (\bibinfo  {publisher} {Springer,
  Berlin},\ \bibinfo {year} {2014})\ pp.\ \bibinfo {pages}
  {301--318}\BibitemShut {NoStop}%
\bibitem [{\citenamefont {Zeiher}\ \emph {et~al.}(2016)\citenamefont {Zeiher},
  \citenamefont {van Bijnen}, \citenamefont {Schau{\ss}}, \citenamefont {Hild},
  \citenamefont {Choi}, \citenamefont {Pohl}, \citenamefont {Bloch},\ and\
  \citenamefont {Gross}}]{Zeiher_etal2016}%
  \BibitemOpen
  \bibfield  {author} {\bibinfo {author} {\bibfnamefont {J.}~\bibnamefont
  {Zeiher}}, \bibinfo {author} {\bibfnamefont {R.}~\bibnamefont {van Bijnen}},
  \bibinfo {author} {\bibfnamefont {P.}~\bibnamefont {Schau{\ss}}}, \bibinfo
  {author} {\bibfnamefont {S.}~\bibnamefont {Hild}}, \bibinfo {author}
  {\bibfnamefont {J.}~\bibnamefont {Choi}}, \bibinfo {author} {\bibfnamefont
  {T.}~\bibnamefont {Pohl}}, \bibinfo {author} {\bibfnamefont {I.}~\bibnamefont
  {Bloch}}, \ and\ \bibinfo {author} {\bibfnamefont {C.}~\bibnamefont
  {Gross}},\ }\bibfield  {title} {\enquote {\bibinfo {title} {Many-body
  interferometry of a {R}ydberg-dressed spin lattice},}\ }\href {\doibase
  10.1038/nphys3835} {\bibfield  {journal} {\bibinfo  {journal} {Nat. Phys.}\
  }\textbf {\bibinfo {volume} {12}},\ \bibinfo {pages} {1095--1099} (\bibinfo
  {year} {2016})}\BibitemShut {NoStop}%
\bibitem [{\citenamefont {Dressel}\ \emph {et~al.}(2018)\citenamefont
  {Dressel}, \citenamefont {Alonso}, \citenamefont {Waegell},\ and\
  \citenamefont {Halpern}}]{Dresseletal18}%
  \BibitemOpen
  \bibfield  {author} {\bibinfo {author} {\bibfnamefont {J.}~\bibnamefont
  {Dressel}}, \bibinfo {author} {\bibfnamefont {J.~R.~G.}\ \bibnamefont
  {Alonso}}, \bibinfo {author} {\bibfnamefont {M.}~\bibnamefont {Waegell}}, \
  and\ \bibinfo {author} {\bibfnamefont {N.~Y.}\ \bibnamefont {Halpern}},\
  }\bibfield  {title} {\enquote {\bibinfo {title} {Strengthening weak
  measurements of qubit out-of-time-order correlators},}\ }\href {\doibase
  10.1103/PhysRevA.98.012132} {\bibfield  {journal} {\bibinfo  {journal} {Phys.
  Rev. A.}\ }\textbf {\bibinfo {volume} {98}},\ \bibinfo {pages} {012132}
  (\bibinfo {year} {2018})}\BibitemShut {NoStop}%
\bibitem [{\citenamefont {Zeiher}\ \emph {et~al.}(2017)\citenamefont {Zeiher},
  \citenamefont {Choi}, \citenamefont {Rubio-Abadal}, \citenamefont {Pohl},
  \citenamefont {{van Bijnen}}, \citenamefont {Bloch},\ and\ \citenamefont
  {Gross}}]{zeiher2017a}%
  \BibitemOpen
  \bibfield  {author} {\bibinfo {author} {\bibfnamefont {J.}~\bibnamefont
  {Zeiher}}, \bibinfo {author} {\bibfnamefont {J.-Y.}\ \bibnamefont {Choi}},
  \bibinfo {author} {\bibfnamefont {A.}~\bibnamefont {Rubio-Abadal}}, \bibinfo
  {author} {\bibfnamefont {T.}~\bibnamefont {Pohl}}, \bibinfo {author}
  {\bibfnamefont {R.}~\bibnamefont {{van Bijnen}}}, \bibinfo {author}
  {\bibfnamefont {I.}~\bibnamefont {Bloch}}, \ and\ \bibinfo {author}
  {\bibfnamefont {C.}~\bibnamefont {Gross}},\ }\bibfield  {title} {\enquote
  {\bibinfo {title} {Coherent many-body spin dynamics in a long-range
  interacting {I}sing chain},}\ }\href {\doibase 10/gcqmz6} {\bibfield
  {journal} {\bibinfo  {journal} {Phys. Rev. X}\ }\textbf {\bibinfo {volume}
  {7}},\ \bibinfo {pages} {041063} (\bibinfo {year} {2017})}\BibitemShut
  {NoStop}%
\bibitem [{\citenamefont {Glaetzle}\ \emph {et~al.}(2015)\citenamefont
  {Glaetzle}, \citenamefont {Dalmonte}, \citenamefont {Nath}, \citenamefont
  {Gross}, \citenamefont {Bloch},\ and\ \citenamefont
  {Zoller}}]{Glaetzle_etal2015}%
  \BibitemOpen
  \bibfield  {author} {\bibinfo {author} {\bibfnamefont {A.~W.}\ \bibnamefont
  {Glaetzle}}, \bibinfo {author} {\bibfnamefont {M.}~\bibnamefont {Dalmonte}},
  \bibinfo {author} {\bibfnamefont {R.}~\bibnamefont {Nath}}, \bibinfo {author}
  {\bibfnamefont {C.}~\bibnamefont {Gross}}, \bibinfo {author} {\bibfnamefont
  {I.}~\bibnamefont {Bloch}}, \ and\ \bibinfo {author} {\bibfnamefont
  {P.}~\bibnamefont {Zoller}},\ }\bibfield  {title} {\enquote {\bibinfo {title}
  {Designing frustrated quantum magnets with laser-dressed {R}ydberg atoms},}\
  }\href {\doibase 10.1103/PhysRevLett.114.173002} {\bibfield  {journal}
  {\bibinfo  {journal} {Phys. Rev. Lett.}\ }\textbf {\bibinfo {volume} {114}},\
  \bibinfo {pages} {173002} (\bibinfo {year} {2015})}\BibitemShut {NoStop}%
\bibitem [{\citenamefont {{van Bijnen}}\ and\ \citenamefont
  {Pohl}(2015)}]{vanbijnen2015}%
  \BibitemOpen
  \bibfield  {author} {\bibinfo {author} {\bibfnamefont {R.~M.~W.}\
  \bibnamefont {{van Bijnen}}}\ and\ \bibinfo {author} {\bibfnamefont
  {T.}~\bibnamefont {Pohl}},\ }\bibfield  {title} {\enquote {\bibinfo {title}
  {Quantum magnetism and topological ordering via {R}ydberg dressing near
  {F}{\"o}rster resonances},}\ }\href {\doibase 10/gddh9d} {\bibfield
  {journal} {\bibinfo  {journal} {Phys. Rev. Lett.}\ }\textbf {\bibinfo
  {volume} {114}},\ \bibinfo {pages} {243002} (\bibinfo {year}
  {2015})}\BibitemShut {NoStop}%
\bibitem [{\citenamefont {Guardado-Sanchez}\ \emph {et~al.}()\citenamefont
  {Guardado-Sanchez}, \citenamefont {Brown}, \citenamefont {Mitra},
  \citenamefont {Devakul}, \citenamefont {Huse}, \citenamefont {Schauss},\ and\
  \citenamefont {Bakr}}]{guardado-sanchez2017}%
  \BibitemOpen
  \bibfield  {author} {\bibinfo {author} {\bibfnamefont {E.}~\bibnamefont
  {Guardado-Sanchez}}, \bibinfo {author} {\bibfnamefont {P.~T.}\ \bibnamefont
  {Brown}}, \bibinfo {author} {\bibfnamefont {D.}~\bibnamefont {Mitra}},
  \bibinfo {author} {\bibfnamefont {T.}~\bibnamefont {Devakul}}, \bibinfo
  {author} {\bibfnamefont {D.~A.}\ \bibnamefont {Huse}}, \bibinfo {author}
  {\bibfnamefont {P.}~\bibnamefont {Schauss}}, \ and\ \bibinfo {author}
  {\bibfnamefont {W.~S.}\ \bibnamefont {Bakr}},\ }\href@noop {} {\enquote
  {\bibinfo {title} {Probing quench dynamics across a quantum phase transition
  into a $2${D} {I}sing antiferromagnet},}\ }\Eprint
  {http://arxiv.org/abs/1711.00887} {arXiv:1711.00887} \BibitemShut {NoStop}%
\bibitem [{\citenamefont {Labuhn}\ \emph {et~al.}(2016)\citenamefont {Labuhn},
  \citenamefont {Barredo}, \citenamefont {Ravets}, \citenamefont {{de
  L{\'e}s{\'e}leuc}}, \citenamefont {Macr{\`\i}}, \citenamefont {Lahaye},\ and\
  \citenamefont {Browaeys}}]{labuhn2016}%
  \BibitemOpen
  \bibfield  {author} {\bibinfo {author} {\bibfnamefont {H.}~\bibnamefont
  {Labuhn}}, \bibinfo {author} {\bibfnamefont {D.}~\bibnamefont {Barredo}},
  \bibinfo {author} {\bibfnamefont {S.}~\bibnamefont {Ravets}}, \bibinfo
  {author} {\bibfnamefont {S.}~\bibnamefont {{de L{\'e}s{\'e}leuc}}}, \bibinfo
  {author} {\bibfnamefont {T.}~\bibnamefont {Macr{\`\i}}}, \bibinfo {author}
  {\bibfnamefont {T.}~\bibnamefont {Lahaye}}, \ and\ \bibinfo {author}
  {\bibfnamefont {A.}~\bibnamefont {Browaeys}},\ }\bibfield  {title} {\enquote
  {\bibinfo {title} {Tunable two-dimensional arrays of single {R}ydberg atoms
  for realizing quantum {I}sing models},}\ }\href {\doibase 10/f8sjhr}
  {\bibfield  {journal} {\bibinfo  {journal} {Nature}\ }\textbf {\bibinfo
  {volume} {534}},\ \bibinfo {pages} {667--670} (\bibinfo {year}
  {2016})}\BibitemShut {NoStop}%
\bibitem [{\citenamefont {Bernien}\ \emph {et~al.}(2017)\citenamefont
  {Bernien}, \citenamefont {Schwartz}, \citenamefont {Keesling}, \citenamefont
  {Levine}, \citenamefont {Omran}, \citenamefont {Pichler}, \citenamefont
  {Choi}, \citenamefont {Zibrov}, \citenamefont {Endres}, \citenamefont
  {Greiner}, \citenamefont {Vuleti{\'c}},\ and\ \citenamefont
  {Lukin}}]{bernien2017}%
  \BibitemOpen
  \bibfield  {author} {\bibinfo {author} {\bibfnamefont {H.}~\bibnamefont
  {Bernien}}, \bibinfo {author} {\bibfnamefont {S.}~\bibnamefont {Schwartz}},
  \bibinfo {author} {\bibfnamefont {A.}~\bibnamefont {Keesling}}, \bibinfo
  {author} {\bibfnamefont {H.}~\bibnamefont {Levine}}, \bibinfo {author}
  {\bibfnamefont {A.}~\bibnamefont {Omran}}, \bibinfo {author} {\bibfnamefont
  {H.}~\bibnamefont {Pichler}}, \bibinfo {author} {\bibfnamefont
  {S.}~\bibnamefont {Choi}}, \bibinfo {author} {\bibfnamefont {A.~S.}\
  \bibnamefont {Zibrov}}, \bibinfo {author} {\bibfnamefont {M.}~\bibnamefont
  {Endres}}, \bibinfo {author} {\bibfnamefont {M.}~\bibnamefont {Greiner}},
  \bibinfo {author} {\bibfnamefont {V.}~\bibnamefont {Vuleti{\'c}}}, \ and\
  \bibinfo {author} {\bibfnamefont {M.~D.}\ \bibnamefont {Lukin}},\ }\bibfield
  {title} {\enquote {\bibinfo {title} {Probing many-body dynamics on a 51-atom
  quantum simulator},}\ }\href {\doibase 10/cgwr} {\bibfield  {journal}
  {\bibinfo  {journal} {Nature}\ }\textbf {\bibinfo {volume} {551}},\ \bibinfo
  {pages} {579--584} (\bibinfo {year} {2017})}\BibitemShut {NoStop}%
\bibitem [{\citenamefont {Fukuhara}\ \emph {et~al.}(2013)\citenamefont
  {Fukuhara}, \citenamefont {Kantian}, \citenamefont {Endres}, \citenamefont
  {Cheneau}, \citenamefont {Schau{\ss}}, \citenamefont {Hild}, \citenamefont
  {Bellem}, \citenamefont {Schollwock}, \citenamefont {Giamarchi},
  \citenamefont {Gross}, \citenamefont {Bloch},\ and\ \citenamefont
  {Kuhr}}]{fukuhara2013a}%
  \BibitemOpen
  \bibfield  {author} {\bibinfo {author} {\bibfnamefont {T.}~\bibnamefont
  {Fukuhara}}, \bibinfo {author} {\bibfnamefont {A.}~\bibnamefont {Kantian}},
  \bibinfo {author} {\bibfnamefont {M.}~\bibnamefont {Endres}}, \bibinfo
  {author} {\bibfnamefont {M.}~\bibnamefont {Cheneau}}, \bibinfo {author}
  {\bibfnamefont {P.}~\bibnamefont {Schau{\ss}}}, \bibinfo {author}
  {\bibfnamefont {S.}~\bibnamefont {Hild}}, \bibinfo {author} {\bibfnamefont
  {D.}~\bibnamefont {Bellem}}, \bibinfo {author} {\bibfnamefont
  {U.}~\bibnamefont {Schollwock}}, \bibinfo {author} {\bibfnamefont
  {T.}~\bibnamefont {Giamarchi}}, \bibinfo {author} {\bibfnamefont
  {C.}~\bibnamefont {Gross}}, \bibinfo {author} {\bibfnamefont
  {I.}~\bibnamefont {Bloch}}, \ and\ \bibinfo {author} {\bibfnamefont
  {S.}~\bibnamefont {Kuhr}},\ }\bibfield  {title} {\enquote {\bibinfo {title}
  {Quantum dynamics of a mobile spin impurity},}\ }\href {\doibase 10/f4s44t}
  {\bibfield  {journal} {\bibinfo  {journal} {Nat. Phys.}\ }\textbf {\bibinfo
  {volume} {9}},\ \bibinfo {pages} {235--241} (\bibinfo {year}
  {2013})}\BibitemShut {NoStop}%
\bibitem [{\citenamefont {Boll}\ \emph {et~al.}(2016)\citenamefont {Boll},
  \citenamefont {Hilker}, \citenamefont {Salomon}, \citenamefont {Omran},
  \citenamefont {Nespolo}, \citenamefont {Pollet}, \citenamefont {Bloch},\ and\
  \citenamefont {Gross}}]{Boll_etal16}%
  \BibitemOpen
  \bibfield  {author} {\bibinfo {author} {\bibfnamefont {M.}~\bibnamefont
  {Boll}}, \bibinfo {author} {\bibfnamefont {T.~A.}\ \bibnamefont {Hilker}},
  \bibinfo {author} {\bibfnamefont {G.}~\bibnamefont {Salomon}}, \bibinfo
  {author} {\bibfnamefont {A.}~\bibnamefont {Omran}}, \bibinfo {author}
  {\bibfnamefont {J.}~\bibnamefont {Nespolo}}, \bibinfo {author} {\bibfnamefont
  {L.}~\bibnamefont {Pollet}}, \bibinfo {author} {\bibfnamefont
  {I.}~\bibnamefont {Bloch}}, \ and\ \bibinfo {author} {\bibfnamefont
  {C.}~\bibnamefont {Gross}},\ }\bibfield  {title} {\enquote {\bibinfo {title}
  {Spin- and density-resolved microscopy of antiferromagnetic correlations in
  {F}ermi-{H}ubbard chains},}\ }\href {\doibase 10.1126/science.aag1635}
  {\bibfield  {journal} {\bibinfo  {journal} {Science}\ }\textbf {\bibinfo
  {volume} {353}},\ \bibinfo {pages} {1257--1260} (\bibinfo {year}
  {2016})}\BibitemShut {NoStop}%
\bibitem [{\citenamefont {Parsons}\ \emph {et~al.}(2016)\citenamefont
  {Parsons}, \citenamefont {Mazurenko}, \citenamefont {Chiu}, \citenamefont
  {Ji}, \citenamefont {Greif},\ and\ \citenamefont {Greiner}}]{Parsons_etal16}%
  \BibitemOpen
  \bibfield  {author} {\bibinfo {author} {\bibfnamefont {M.~F.}\ \bibnamefont
  {Parsons}}, \bibinfo {author} {\bibfnamefont {A.}~\bibnamefont {Mazurenko}},
  \bibinfo {author} {\bibfnamefont {C.~S.}\ \bibnamefont {Chiu}}, \bibinfo
  {author} {\bibfnamefont {G.}~\bibnamefont {Ji}}, \bibinfo {author}
  {\bibfnamefont {D.}~\bibnamefont {Greif}}, \ and\ \bibinfo {author}
  {\bibfnamefont {M.}~\bibnamefont {Greiner}},\ }\bibfield  {title} {\enquote
  {\bibinfo {title} {Site-resolved measurement of the spin-correlation function
  in the {F}ermi-{H}ubbard model},}\ }\href {\doibase 10.1126/science.aag1430}
  {\bibfield  {journal} {\bibinfo  {journal} {Science}\ }\textbf {\bibinfo
  {volume} {353}},\ \bibinfo {pages} {1253--1256} (\bibinfo {year}
  {2016})}\BibitemShut {NoStop}%
\bibitem [{\citenamefont {Cheuk}\ \emph {et~al.}(2016)\citenamefont {Cheuk},
  \citenamefont {Nichols}, \citenamefont {Lawrence}, \citenamefont {Okan},
  \citenamefont {Zhang}, \citenamefont {Khatami}, \citenamefont {Trivedi},
  \citenamefont {Paiva}, \citenamefont {Rigol},\ and\ \citenamefont
  {Zwierlein}}]{cheuk2016a}%
  \BibitemOpen
  \bibfield  {author} {\bibinfo {author} {\bibfnamefont {L.~W.}\ \bibnamefont
  {Cheuk}}, \bibinfo {author} {\bibfnamefont {M.~A.}\ \bibnamefont {Nichols}},
  \bibinfo {author} {\bibfnamefont {K.~R.}\ \bibnamefont {Lawrence}}, \bibinfo
  {author} {\bibfnamefont {M.}~\bibnamefont {Okan}}, \bibinfo {author}
  {\bibfnamefont {H.}~\bibnamefont {Zhang}}, \bibinfo {author} {\bibfnamefont
  {E.}~\bibnamefont {Khatami}}, \bibinfo {author} {\bibfnamefont
  {N.}~\bibnamefont {Trivedi}}, \bibinfo {author} {\bibfnamefont
  {T.}~\bibnamefont {Paiva}}, \bibinfo {author} {\bibfnamefont
  {M.}~\bibnamefont {Rigol}}, \ and\ \bibinfo {author} {\bibfnamefont {M.~W.}\
  \bibnamefont {Zwierlein}},\ }\bibfield  {title} {\enquote {\bibinfo {title}
  {Observation of spatial charge and spin correlations in the $2${D}
  {F}ermi-{H}ubbard model},}\ }\href {\doibase 10/f84js8} {\bibfield  {journal}
  {\bibinfo  {journal} {Science}\ }\textbf {\bibinfo {volume} {353}},\ \bibinfo
  {pages} {1260--1264} (\bibinfo {year} {2016})}\BibitemShut {NoStop}%
\bibitem [{\citenamefont {Brown}\ \emph {et~al.}(2017)\citenamefont {Brown},
  \citenamefont {Mitra}, \citenamefont {Guardado-Sanchez}, \citenamefont
  {Schau{\ss}}, \citenamefont {Kondov}, \citenamefont {Khatami}, \citenamefont
  {Paiva}, \citenamefont {Trivedi}, \citenamefont {Huse},\ and\ \citenamefont
  {Bakr}}]{brown2017}%
  \BibitemOpen
  \bibfield  {author} {\bibinfo {author} {\bibfnamefont {P.~T.}\ \bibnamefont
  {Brown}}, \bibinfo {author} {\bibfnamefont {D.}~\bibnamefont {Mitra}},
  \bibinfo {author} {\bibfnamefont {E.}~\bibnamefont {Guardado-Sanchez}},
  \bibinfo {author} {\bibfnamefont {P.}~\bibnamefont {Schau{\ss}}}, \bibinfo
  {author} {\bibfnamefont {S.~S.}\ \bibnamefont {Kondov}}, \bibinfo {author}
  {\bibfnamefont {E.}~\bibnamefont {Khatami}}, \bibinfo {author} {\bibfnamefont
  {T.}~\bibnamefont {Paiva}}, \bibinfo {author} {\bibfnamefont
  {N.}~\bibnamefont {Trivedi}}, \bibinfo {author} {\bibfnamefont {D.~A.}\
  \bibnamefont {Huse}}, \ and\ \bibinfo {author} {\bibfnamefont {W.~S.}\
  \bibnamefont {Bakr}},\ }\bibfield  {title} {\enquote {\bibinfo {title}
  {Spin-imbalance in a $2${D} {F}ermi-{H}ubbard system},}\ }\href {\doibase
  10/gb26js} {\bibfield  {journal} {\bibinfo  {journal} {Science}\ }\textbf
  {\bibinfo {volume} {357}},\ \bibinfo {pages} {1385--1388} (\bibinfo {year}
  {2017})}\BibitemShut {NoStop}%
\bibitem [{\citenamefont {Cooper}\ \emph {et~al.}()\citenamefont {Cooper},
  \citenamefont {Covey}, \citenamefont {Madjarov}, \citenamefont {Porsev},
  \citenamefont {Safronova},\ and\ \citenamefont {Endres}}]{cooper2018b}%
  \BibitemOpen
  \bibfield  {author} {\bibinfo {author} {\bibfnamefont {A.}~\bibnamefont
  {Cooper}}, \bibinfo {author} {\bibfnamefont {J.~P.}\ \bibnamefont {Covey}},
  \bibinfo {author} {\bibfnamefont {I.~S.}\ \bibnamefont {Madjarov}}, \bibinfo
  {author} {\bibfnamefont {S.~G.}\ \bibnamefont {Porsev}}, \bibinfo {author}
  {\bibfnamefont {M.~S.}\ \bibnamefont {Safronova}}, \ and\ \bibinfo {author}
  {\bibfnamefont {M.}~\bibnamefont {Endres}},\ }\href@noop {} {\enquote
  {\bibinfo {title} {Alkaline earth atoms in optical tweezers},}\ }\Eprint
  {http://arxiv.org/abs/1810.06537} {arXiv:1810.06537} \BibitemShut {NoStop}%
\bibitem [{\citenamefont {Norcia}\ \emph {et~al.}()\citenamefont {Norcia},
  \citenamefont {Young},\ and\ \citenamefont {Kaufman}}]{Norcia_etal}%
  \BibitemOpen
  \bibfield  {author} {\bibinfo {author} {\bibfnamefont {M.~A.}\ \bibnamefont
  {Norcia}}, \bibinfo {author} {\bibfnamefont {A.~W.}\ \bibnamefont {Young}}, \
  and\ \bibinfo {author} {\bibfnamefont {A.~M.}\ \bibnamefont {Kaufman}},\
  }\href@noop {} {\enquote {\bibinfo {title} {Microscopic control and detection
  of ultracold strontium in optical-tweezer arrays},}\ }\Eprint
  {http://arxiv.org/abs/1810.06626} {arXiv:1810.06626} \BibitemShut {NoStop}%
\bibitem [{\citenamefont {Nagourney}\ \emph {et~al.}(1986)\citenamefont
  {Nagourney}, \citenamefont {Sandberg},\ and\ \citenamefont
  {Dehmelt}}]{nagourney1986}%
  \BibitemOpen
  \bibfield  {author} {\bibinfo {author} {\bibfnamefont {W.}~\bibnamefont
  {Nagourney}}, \bibinfo {author} {\bibfnamefont {J.}~\bibnamefont {Sandberg}},
  \ and\ \bibinfo {author} {\bibfnamefont {H.}~\bibnamefont {Dehmelt}},\
  }\bibfield  {title} {\enquote {\bibinfo {title} {Shelved optical electron
  amplifier: {O}bservation of quantum jumps},}\ }\href {\doibase 10/fwtw5p}
  {\bibfield  {journal} {\bibinfo  {journal} {Phys. Rev. Lett.}\ }\textbf
  {\bibinfo {volume} {56}},\ \bibinfo {pages} {2797--2799} (\bibinfo {year}
  {1986})}\BibitemShut {NoStop}%
\bibitem [{\citenamefont {Sauter}\ \emph {et~al.}(1986)\citenamefont {Sauter},
  \citenamefont {Neuhauser}, \citenamefont {Blatt},\ and\ \citenamefont
  {Toschek}}]{sauter1986}%
  \BibitemOpen
  \bibfield  {author} {\bibinfo {author} {\bibfnamefont {T.}~\bibnamefont
  {Sauter}}, \bibinfo {author} {\bibfnamefont {W.}~\bibnamefont {Neuhauser}},
  \bibinfo {author} {\bibfnamefont {R.}~\bibnamefont {Blatt}}, \ and\ \bibinfo
  {author} {\bibfnamefont {P.~E.}\ \bibnamefont {Toschek}},\ }\bibfield
  {title} {\enquote {\bibinfo {title} {Observation of quantum jumps},}\ }\href
  {\doibase 10/dzkmrh} {\bibfield  {journal} {\bibinfo  {journal} {Phys. Rev.
  Lett.}\ }\textbf {\bibinfo {volume} {57}},\ \bibinfo {pages} {1696--1698}
  (\bibinfo {year} {1986})}\BibitemShut {NoStop}%
\bibitem [{\citenamefont {Bergquist}\ \emph {et~al.}(1986)\citenamefont
  {Bergquist}, \citenamefont {Hulet}, \citenamefont {Itano},\ and\
  \citenamefont {Wineland}}]{bergquist1986}%
  \BibitemOpen
  \bibfield  {author} {\bibinfo {author} {\bibfnamefont {J.~C.}\ \bibnamefont
  {Bergquist}}, \bibinfo {author} {\bibfnamefont {Randall~G.}\ \bibnamefont
  {Hulet}}, \bibinfo {author} {\bibfnamefont {Wayne~M.}\ \bibnamefont {Itano}},
  \ and\ \bibinfo {author} {\bibfnamefont {D.~J.}\ \bibnamefont {Wineland}},\
  }\bibfield  {title} {\enquote {\bibinfo {title} {Observation of quantum jumps
  in a single atom},}\ }\href {\doibase 10/cvcqss} {\bibfield  {journal}
  {\bibinfo  {journal} {Phys. Rev. Lett.}\ }\textbf {\bibinfo {volume} {57}},\
  \bibinfo {pages} {1699--1702} (\bibinfo {year} {1986})}\BibitemShut {NoStop}%
\bibitem [{\citenamefont {Barredo}\ \emph {et~al.}(2016)\citenamefont
  {Barredo}, \citenamefont {de~L{\'e}s{\'e}leuc}, \citenamefont {Lienhard},
  \citenamefont {Lahaye},\ and\ \citenamefont {Browaeys}}]{Barredo_etal16}%
  \BibitemOpen
  \bibfield  {author} {\bibinfo {author} {\bibfnamefont {D.}~\bibnamefont
  {Barredo}}, \bibinfo {author} {\bibfnamefont {S.}~\bibnamefont
  {de~L{\'e}s{\'e}leuc}}, \bibinfo {author} {\bibfnamefont {V.}~\bibnamefont
  {Lienhard}}, \bibinfo {author} {\bibfnamefont {T.}~\bibnamefont {Lahaye}}, \
  and\ \bibinfo {author} {\bibfnamefont {A.}~\bibnamefont {Browaeys}},\
  }\bibfield  {title} {\enquote {\bibinfo {title} {An atom-by-atom assembler of
  defect-free arbitrary two-dimensional atomic arrays},}\ }\href {\doibase
  10.1126/science.aah3778} {\bibfield  {journal} {\bibinfo  {journal}
  {Science}\ }\textbf {\bibinfo {volume} {354}},\ \bibinfo {pages} {1021--1023}
  (\bibinfo {year} {2016})}\BibitemShut {NoStop}%
\bibitem [{\citenamefont {Endres}\ \emph {et~al.}(2016)\citenamefont {Endres},
  \citenamefont {Bernien}, \citenamefont {Keesling}, \citenamefont {Levine},
  \citenamefont {Anschuetz}, \citenamefont {Krajenbrink}, \citenamefont
  {Senko}, \citenamefont {Vuletic}, \citenamefont {Greiner},\ and\
  \citenamefont {Lukin}}]{endres2016}%
  \BibitemOpen
  \bibfield  {author} {\bibinfo {author} {\bibfnamefont {M.}~\bibnamefont
  {Endres}}, \bibinfo {author} {\bibfnamefont {H.}~\bibnamefont {Bernien}},
  \bibinfo {author} {\bibfnamefont {A.}~\bibnamefont {Keesling}}, \bibinfo
  {author} {\bibfnamefont {H.}~\bibnamefont {Levine}}, \bibinfo {author}
  {\bibfnamefont {E.~R.}\ \bibnamefont {Anschuetz}}, \bibinfo {author}
  {\bibfnamefont {A.}~\bibnamefont {Krajenbrink}}, \bibinfo {author}
  {\bibfnamefont {C.}~\bibnamefont {Senko}}, \bibinfo {author} {\bibfnamefont
  {V.}~\bibnamefont {Vuletic}}, \bibinfo {author} {\bibfnamefont
  {M.}~\bibnamefont {Greiner}}, \ and\ \bibinfo {author} {\bibfnamefont
  {M.~D.}\ \bibnamefont {Lukin}},\ }\bibfield  {title} {\enquote {\bibinfo
  {title} {Atom-by-atom assembly of defect-free one-dimensional cold atom
  arrays},}\ }\href {\doibase 10/bs3c} {\bibfield  {journal} {\bibinfo
  {journal} {Science}\ }\textbf {\bibinfo {volume} {354}},\ \bibinfo {pages}
  {1024--1027} (\bibinfo {year} {2016})}\BibitemShut {NoStop}%
\bibitem [{\citenamefont {Endres}\ \emph {et~al.}(2013)\citenamefont {Endres},
  \citenamefont {Cheneau}, \citenamefont {Fukuhara}, \citenamefont
  {Weitenberg}, \citenamefont {Schau{\ss}}, \citenamefont {Gross},
  \citenamefont {Mazza}, \citenamefont {Ba{\~n}uls}, \citenamefont {Pollet},
  \citenamefont {Bloch},\ and\ \citenamefont {Kuhr}}]{endres2013}%
  \BibitemOpen
  \bibfield  {author} {\bibinfo {author} {\bibfnamefont {M.}~\bibnamefont
  {Endres}}, \bibinfo {author} {\bibfnamefont {M.}~\bibnamefont {Cheneau}},
  \bibinfo {author} {\bibfnamefont {T.}~\bibnamefont {Fukuhara}}, \bibinfo
  {author} {\bibfnamefont {C.}~\bibnamefont {Weitenberg}}, \bibinfo {author}
  {\bibfnamefont {P.}~\bibnamefont {Schau{\ss}}}, \bibinfo {author}
  {\bibfnamefont {C.}~\bibnamefont {Gross}}, \bibinfo {author} {\bibfnamefont
  {L.}~\bibnamefont {Mazza}}, \bibinfo {author} {\bibfnamefont
  {M.}~\bibnamefont {Ba{\~n}uls}}, \bibinfo {author} {\bibfnamefont
  {L.}~\bibnamefont {Pollet}}, \bibinfo {author} {\bibfnamefont
  {I.}~\bibnamefont {Bloch}}, \ and\ \bibinfo {author} {\bibfnamefont
  {S.}~\bibnamefont {Kuhr}},\ }\bibfield  {title} {\enquote {\bibinfo {title}
  {Single-site- and single-atom-resolved measurement of correlation
  functions},}\ }\href {\doibase 10/gddjfz} {\bibfield  {journal} {\bibinfo
  {journal} {Appl. Phys. B}\ }\textbf {\bibinfo {volume} {113}},\ \bibinfo
  {pages} {27} (\bibinfo {year} {2013})}\BibitemShut {NoStop}%
\bibitem [{\citenamefont {Lester}\ \emph {et~al.}(2014)\citenamefont {Lester},
  \citenamefont {Kaufman},\ and\ \citenamefont {Regal}}]{lester2014}%
  \BibitemOpen
  \bibfield  {author} {\bibinfo {author} {\bibfnamefont {B.~J.}\ \bibnamefont
  {Lester}}, \bibinfo {author} {\bibfnamefont {A.~M.}\ \bibnamefont {Kaufman}},
  \ and\ \bibinfo {author} {\bibfnamefont {C.~A.}\ \bibnamefont {Regal}},\
  }\bibfield  {title} {\enquote {\bibinfo {title} {Raman cooling imaging:
  {D}etecting single atoms near their ground state of motion},}\ }\href
  {\doibase 10/gddjqz} {\bibfield  {journal} {\bibinfo  {journal} {Phys. Rev.
  A}\ }\textbf {\bibinfo {volume} {90}},\ \bibinfo {pages} {011804} (\bibinfo
  {year} {2014})}\BibitemShut {NoStop}%
\bibitem [{\citenamefont {Barredo}\ \emph {et~al.}(2018)\citenamefont
  {Barredo}, \citenamefont {Lienhard}, \citenamefont {{de L{\'e}s{\'e}leuc}},
  \citenamefont {Lahaye},\ and\ \citenamefont {Browaeys}}]{barredo2017a}%
  \BibitemOpen
  \bibfield  {author} {\bibinfo {author} {\bibfnamefont {D.}~\bibnamefont
  {Barredo}}, \bibinfo {author} {\bibfnamefont {V.}~\bibnamefont {Lienhard}},
  \bibinfo {author} {\bibfnamefont {S.}~\bibnamefont {{de L{\'e}s{\'e}leuc}}},
  \bibinfo {author} {\bibfnamefont {T.}~\bibnamefont {Lahaye}}, \ and\ \bibinfo
  {author} {\bibfnamefont {A.}~\bibnamefont {Browaeys}},\ }\bibfield  {title}
  {\enquote {\bibinfo {title} {Synthetic three-dimensional atomic structures
  assembled atom by atom},}\ }\href {\doibase 10.1038/s41586-018-0450-2}
  {\bibfield  {journal} {\bibinfo  {journal} {Nature}\ }\textbf {\bibinfo
  {volume} {561}},\ \bibinfo {pages} {79--82} (\bibinfo {year}
  {2018})}\BibitemShut {NoStop}%
\bibitem [{\citenamefont {Marchioro}\ \emph {et~al.}(1978)\citenamefont
  {Marchioro}, \citenamefont {Pellegrinotti}, \citenamefont {Pulvirenti},\ and\
  \citenamefont {Triolo}}]{Marchioro_etal78}%
  \BibitemOpen
  \bibfield  {author} {\bibinfo {author} {\bibfnamefont {C.}~\bibnamefont
  {Marchioro}}, \bibinfo {author} {\bibfnamefont {A.}~\bibnamefont
  {Pellegrinotti}}, \bibinfo {author} {\bibfnamefont {M.}~\bibnamefont
  {Pulvirenti}}, \ and\ \bibinfo {author} {\bibfnamefont {L.}~\bibnamefont
  {Triolo}},\ }\bibfield  {title} {\enquote {\bibinfo {title} {Velocity of a
  perturbation in infinite lattice systems},}\ }\href {\doibase
  10.1007/BF01011695} {\bibfield  {journal} {\bibinfo  {journal} {J. Stat.
  Phys.}\ }\textbf {\bibinfo {volume} {19}},\ \bibinfo {pages} {499--510}
  (\bibinfo {year} {1978})}\BibitemShut {NoStop}%
\bibitem [{\citenamefont {Hastings}\ and\ \citenamefont
  {Koma}(2006)}]{HastingsKoma06}%
  \BibitemOpen
  \bibfield  {author} {\bibinfo {author} {\bibfnamefont {M.~B.}\ \bibnamefont
  {Hastings}}\ and\ \bibinfo {author} {\bibfnamefont {T.}~\bibnamefont
  {Koma}},\ }\bibfield  {title} {\enquote {\bibinfo {title} {Spectral gap and
  exponential decay of correlations},}\ }\href {\doibase
  10.1007/s00220-006-0030-4} {\bibfield  {journal} {\bibinfo  {journal}
  {Commun. Math. Phys.}\ }\textbf {\bibinfo {volume} {265}},\ \bibinfo {pages}
  {781--804} (\bibinfo {year} {2006})}\BibitemShut {NoStop}%
\bibitem [{\citenamefont {Nachtergaele}\ and\ \citenamefont
  {Sims}(2006)}]{NachtergaeleSims06}%
  \BibitemOpen
  \bibfield  {author} {\bibinfo {author} {\bibfnamefont {B.}~\bibnamefont
  {Nachtergaele}}\ and\ \bibinfo {author} {\bibfnamefont {R.}~\bibnamefont
  {Sims}},\ }\bibfield  {title} {\enquote {\bibinfo {title} {{L}ieb-{R}obinson
  bounds and the exponential clustering theorem},}\ }\href {\doibase
  10.1007/s00220-006-1556-1} {\bibfield  {journal} {\bibinfo  {journal}
  {Commun. Math. Phys.}\ }\textbf {\bibinfo {volume} {265}},\ \bibinfo {pages}
  {119--130} (\bibinfo {year} {2006})}\BibitemShut {NoStop}%
\bibitem [{\citenamefont {Burrell}\ and\ \citenamefont
  {Osborne}(2007)}]{BurrellOsborne07}%
  \BibitemOpen
  \bibfield  {author} {\bibinfo {author} {\bibfnamefont {C.~K.}\ \bibnamefont
  {Burrell}}\ and\ \bibinfo {author} {\bibfnamefont {T.~J.}\ \bibnamefont
  {Osborne}},\ }\bibfield  {title} {\enquote {\bibinfo {title} {Bounds on the
  speed of information propagation in disordered quantum spin chains},}\ }\href
  {\doibase 10.1103/PhysRevLett.99.167201} {\bibfield  {journal} {\bibinfo
  {journal} {Phys. Rev. Lett.}\ }\textbf {\bibinfo {volume} {99}},\ \bibinfo
  {pages} {167201} (\bibinfo {year} {2007})}\BibitemShut {NoStop}%
\bibitem [{\citenamefont {Poulin}(2010)}]{Poulin10}%
  \BibitemOpen
  \bibfield  {author} {\bibinfo {author} {\bibfnamefont {D.}~\bibnamefont
  {Poulin}},\ }\bibfield  {title} {\enquote {\bibinfo {title}
  {{L}ieb-{R}obinson bound and locality for general {M}arkovian quantum
  dynamics},}\ }\href {\doibase 10.1103/PhysRevLett.104.190401} {\bibfield
  {journal} {\bibinfo  {journal} {Phys. Rev. Lett.}\ }\textbf {\bibinfo
  {volume} {104}},\ \bibinfo {pages} {190401} (\bibinfo {year}
  {2010})}\BibitemShut {NoStop}%
\bibitem [{\citenamefont {M{\'e}tivier}\ \emph {et~al.}(2014)\citenamefont
  {M{\'e}tivier}, \citenamefont {Bachelard},\ and\ \citenamefont
  {Kastner}}]{MetivierBachelardKastner14}%
  \BibitemOpen
  \bibfield  {author} {\bibinfo {author} {\bibfnamefont {D.}~\bibnamefont
  {M{\'e}tivier}}, \bibinfo {author} {\bibfnamefont {R.}~\bibnamefont
  {Bachelard}}, \ and\ \bibinfo {author} {\bibfnamefont {M.}~\bibnamefont
  {Kastner}},\ }\bibfield  {title} {\enquote {\bibinfo {title} {Spreading of
  perturbations in long-range interacting classical lattice models},}\ }\href
  {\doibase 10.1103/PhysRevLett.112.210601} {\bibfield  {journal} {\bibinfo
  {journal} {Phys. Rev. Lett.}\ }\textbf {\bibinfo {volume} {112}},\ \bibinfo
  {pages} {210601} (\bibinfo {year} {2014})}\BibitemShut {NoStop}%
\bibitem [{\citenamefont {Foss-Feig}\ \emph {et~al.}(2015)\citenamefont
  {Foss-Feig}, \citenamefont {Gong}, \citenamefont {Clark},\ and\ \citenamefont
  {Gorshkov}}]{FossFeigGongClarkGorshkov15}%
  \BibitemOpen
  \bibfield  {author} {\bibinfo {author} {\bibfnamefont {M.}~\bibnamefont
  {Foss-Feig}}, \bibinfo {author} {\bibfnamefont {Z.-X.}\ \bibnamefont {Gong}},
  \bibinfo {author} {\bibfnamefont {C.~W.}\ \bibnamefont {Clark}}, \ and\
  \bibinfo {author} {\bibfnamefont {A.~V.}\ \bibnamefont {Gorshkov}},\
  }\bibfield  {title} {\enquote {\bibinfo {title} {Nearly linear light cones in
  long-range interacting quantum systems},}\ }\href {\doibase
  10.1103/PhysRevLett.114.157201} {\bibfield  {journal} {\bibinfo  {journal}
  {Phys. Rev. Lett.}\ }\textbf {\bibinfo {volume} {114}},\ \bibinfo {pages}
  {157201} (\bibinfo {year} {2015})}\BibitemShut {NoStop}%
\bibitem [{\citenamefont {Storch}\ \emph {et~al.}(2015)\citenamefont {Storch},
  \citenamefont {{van den Worm}},\ and\ \citenamefont
  {Kastner}}]{StorchvandenWormKastner15}%
  \BibitemOpen
  \bibfield  {author} {\bibinfo {author} {\bibfnamefont {D.-M.}\ \bibnamefont
  {Storch}}, \bibinfo {author} {\bibfnamefont {M.}~\bibnamefont {{van den
  Worm}}}, \ and\ \bibinfo {author} {\bibfnamefont {M.}~\bibnamefont
  {Kastner}},\ }\bibfield  {title} {\enquote {\bibinfo {title} {Interplay of
  soundcone and supersonic propagation in lattice models with power law
  interactions},}\ }\href {\doibase 10.1088/1367-2630/17/6/063021} {\bibfield
  {journal} {\bibinfo  {journal} {New J. Phys.}\ }\textbf {\bibinfo {volume}
  {17}},\ \bibinfo {pages} {063021} (\bibinfo {year} {2015})}\BibitemShut
  {NoStop}%
\bibitem [{\citenamefont {Matsuta}\ \emph {et~al.}(2016)\citenamefont
  {Matsuta}, \citenamefont {Koma},\ and\ \citenamefont
  {Nakamura}}]{Matsuta2016}%
  \BibitemOpen
  \bibfield  {author} {\bibinfo {author} {\bibfnamefont {T.}~\bibnamefont
  {Matsuta}}, \bibinfo {author} {\bibfnamefont {T.}~\bibnamefont {Koma}}, \
  and\ \bibinfo {author} {\bibfnamefont {S.}~\bibnamefont {Nakamura}},\
  }\bibfield  {title} {\enquote {\bibinfo {title} {Improving the
  {L}ieb-{R}obinson bound for long-range interactions},}\ }\href {\doibase
  10.1007/s00023-016-0526-1} {\bibfield  {journal} {\bibinfo  {journal} {Ann.
  Henri Poincar\'e}\ }\textbf {\bibinfo {volume} {18}},\ \bibinfo {pages}
  {1--10} (\bibinfo {year} {2016})}\BibitemShut {NoStop}%
\bibitem [{\citenamefont {Abdul-Rahman}\ \emph {et~al.}(2017)\citenamefont
  {Abdul-Rahman}, \citenamefont {Nachtergaele}, \citenamefont {Sims},\ and\
  \citenamefont {Stolz}}]{AbdulRahman_etal17}%
  \BibitemOpen
  \bibfield  {author} {\bibinfo {author} {\bibfnamefont {H.}~\bibnamefont
  {Abdul-Rahman}}, \bibinfo {author} {\bibfnamefont {B.}~\bibnamefont
  {Nachtergaele}}, \bibinfo {author} {\bibfnamefont {R.}~\bibnamefont {Sims}},
  \ and\ \bibinfo {author} {\bibfnamefont {G.}~\bibnamefont {Stolz}},\
  }\bibfield  {title} {\enquote {\bibinfo {title} {Localization properties of
  the disordered {XY} spin chain},}\ }\href {\doibase 10.1002/andp.201600280}
  {\bibfield  {journal} {\bibinfo  {journal} {Ann. Phys. (Berl.)}\ }\textbf
  {\bibinfo {volume} {529}},\ \bibinfo {pages} {1600280} (\bibinfo {year}
  {2017})}\BibitemShut {NoStop}%
\bibitem [{\citenamefont {Kastner}(2015)}]{Kastner2015}%
  \BibitemOpen
  \bibfield  {author} {\bibinfo {author} {\bibfnamefont {M.}~\bibnamefont
  {Kastner}},\ }\bibfield  {title} {\enquote {\bibinfo {title}
  {Entanglement-enhanced spreading of correlations},}\ }\href {\doibase
  10.1088/1367-2630/17/12/123024} {\bibfield  {journal} {\bibinfo  {journal}
  {New J. Phys.}\ }\textbf {\bibinfo {volume} {17}},\ \bibinfo {pages} {123024}
  (\bibinfo {year} {2015})}\BibitemShut {NoStop}%
\bibitem [{\citenamefont {Kaufman}\ \emph {et~al.}(2012)\citenamefont
  {Kaufman}, \citenamefont {Lester},\ and\ \citenamefont
  {Regal}}]{kaufman2012}%
  \BibitemOpen
  \bibfield  {author} {\bibinfo {author} {\bibfnamefont {A.~M.}\ \bibnamefont
  {Kaufman}}, \bibinfo {author} {\bibfnamefont {B.~J.}\ \bibnamefont {Lester}},
  \ and\ \bibinfo {author} {\bibfnamefont {C.~A.}\ \bibnamefont {Regal}},\
  }\bibfield  {title} {\enquote {\bibinfo {title} {Cooling a single atom in an
  optical tweezer to its quantum ground state},}\ }\href {\doibase 10/gddjnd}
  {\bibfield  {journal} {\bibinfo  {journal} {Phys. Rev. X}\ }\textbf {\bibinfo
  {volume} {2}},\ \bibinfo {pages} {041014} (\bibinfo {year}
  {2012})}\BibitemShut {NoStop}%
\end{thebibliography}%
	%\bibliography{/home/philipp/Documents/MendeleyBib/MyPub-modPMP}
	%
\end{document}